\documentclass[ShortAfour,doublespace,sageh]{sagej}
\usepackage{graphicx}
\usepackage{etex}
\usepackage{color}
\usepackage{morefloats}
\usepackage[T1]{fontenc}
\usepackage{calc}
\usepackage{bm}
\usepackage{amsmath}
\usepackage{amssymb}
\usepackage{makeidx}
\usepackage{placeins}
\usepackage{psfrag}
\usepackage{lineno}
\usepackage{grfext}
\usepackage{colortbl}
\usepackage{listings}
\usepackage[font=bf,skip=\baselineskip]{caption}
\usepackage[colorinlistoftodos,backgroundcolor=yellow]{todonotes}
\usepackage{float}
\usepackage{lscape}

\usepackage{moreverb,url}

\usepackage[colorlinks,bookmarksopen,bookmarksnumbered,citecolor=red,urlcolor=red]{hyperref}

\newcommand\BibTeX{{\rmfamily B\kern-.05em \textsc{i\kern-.025em b}\kern-.08em
T\kern-.1667em\lower.7ex\hbox{E}\kern-.125emX}}

\def\volumeyear{2016}
\def\journalname{The International Journal of High Performance Computing Applications}

\newfloat{code}{thp}{lop}
\floatname{code}{Code example}

\DeclareGraphicsExtensions{.pdf}

\graphicspath{{./}}

\newlength{\singleSize} 
\newlength{\tripleSize} 
\definecolor{lightgray}{gray}{0.95}
\definecolor{mediumgray}{gray}{0.8}
\definecolor{black}{gray}{0.0}
\definecolor{mygreen}{rgb}{0,0.6,0}

\DeclareCaptionStyle{listing} [justification=raggedright]{}

\newcounter{nalg} 
\renewcommand{\thenalg}{\arabic{nalg}} 
\DeclareCaptionLabelFormat{algocaption}{Algorithm \thenalg} 

\lstnewenvironment{algorithm}[1][] 
{   
    \refstepcounter{nalg} 
    \lstset{ 
        frame=tB,
        numbers=left, 
        basicstyle=\footnotesize, 
	    columns=fullflexible,
        keywordstyle=\color{black}\bfseries\em,
        keywords={if, else, while, begin, end, } 
        numbers=left,
        #1 
    }
}
{}

\newcommand{\vek}[1]{\boldsymbol{#1}}
\newcommand{\matr}[1]{\text{\textbf{#1}}}

\newcommand{\D}{\text{d}}
\newcommand{\dt}{\text{dt}}

\begin{document}

\runninghead{M\"uller et al.}



\title{Strong scaling for numerical weather prediction at peta\-scale with the atmos\-pheric model NUMA}



\author{Andreas M\"uller\affilnum{1}, Michal A. Kopera\affilnum{1}, Simone Marras\affilnum{2}, Lucas C. Wilcox\affilnum{1}, Tobin Isaac\affilnum{3} and Francis X. Giraldo\affilnum{1}}

\affiliation{\affilnum{1}Department of Applied Mathematics, Naval Postgraduate School, Monterey, CA USA\\
\affilnum{2}Department of Geophysics, Stanford University, Stanford, CA USA\\
\affilnum{3}Computing Institute, University of Chicago, IL, USA}

\corrauth{Andreas M\"uller, work was done at the Naval Postgraduate School, Department of Applied Mathematics, Monterey, CA 93943, USA. Current address: European Center for Medium-Range Weather Forecasts, Shinfield Park, Reading RG2 9AX, UK.}

\email{amueller@anmr.de}

%
%
%

\begin{abstract}

Numerical weather prediction (NWP) has proven to be computationally challenging due to its inherent multiscale nature. Currently, the highest resolution NWP models use a horizontal resolution of about 10$\,$km. At this resolution many important processes in the atmosphere are not resolved. Needless to say this introduces errors. In order to increase the resolution of NWP models highly scalable atmospheric models are needed.

The Non-hydrostatic Unified Model of the Atmosphere (NUMA), developed by the authors at the Naval Postgraduate School, was designed to achieve this purpose. NUMA is used by the Naval Research Laboratory, Monterey as the engine inside its next generation weather prediction system NEPTUNE. NUMA solves the fully compressible Navier-Stokes equations by means of high-order Galerkin methods (both spectral element as well as discontinuous Galerkin methods can be used). Mesh generation is done using the p4est library. NUMA is capable of running middle and upper atmosphere simulations since it does not make use of the shallow-atmosphere approximation.

This paper presents the performance analysis and optimization of the spectral element version of NUMA. The performance at different optimization stages is analyzed using a theoretical performance model as well as measurements via hardware counters. Machine independent optimization is compared to machine specific optimization using BG/Q vector intrinsics. By using vector intrinsics the main computations reach 1.2 PFlops on the entire machine Mira (12\% of the theoretical peak performance). The paper also presents scalability studies for two idealized test cases that are relevant for NWP applications. The atmospheric model NUMA delivers an excellent strong scaling efficiency of 99\% on the entire supercomputer Mira using a mesh with 1.8 billion grid points. This allows to run a global forecast of a baroclinic wave test case at 3$\,$km uniform horizontal resolution and double precision within the time frame required for operational weather prediction.

\end{abstract}


\maketitle




\section{Introduction}

Numerical weather prediction (NWP) has always been considered one of the important computationally intensive uses of supercomputers. Nevertheless there is a big gap between the size of the available supercomputers and the amount of computing power that is used for operational weather prediction. State of the art operational deterministic weather forecasts typically use about 1000 processors \citep{bauer2015quiet} with a global resolution of about 10$\,$km, whereas the biggest available supercomputers offer more than one million processors allowing more than $10^{15}$ floating point operations in one second (petascale). One of the reasons for this discrepancy is that many weather models do not scale to this large number of processors and therefore are not able to make good use of these big machines. The National Oceanic and Atmospheric Administration (NOAA) has initiated the High-Impact Weather Prediction Project (HIWPP) with the goal to reach numerical weather prediction at 3$\,$km resolution by the year 2020 \citep{hiwppProjectPlan}. Being able to improve the resolution by almost one order of magnitude will allow resolving some of the atmospheric processes explicitly that are currently only described by heuristic approximations (parameterizations). For this reason, it is expected that such a significant improvement in the resolution of weather prediction models will reduce the error and improve the accuracy of weather forecasts significantly.

In this paper, we show that the Non-hydrostatic Unified Model of the Atmosphere, NUMA \citep{Giraldo2008a,Kelly2012,giraldo2013implicit}, is capable of simulating a global baroclinic wave test case within the timeframe required for operational weather prediction at 3$\,$km resolution using a uniform global mesh with 31 layers in the vertical direction. We achieve this performance with double precision and without making use of the commonly used shallow atmosphere approximation; in fact, NUMA (within the NEPTUNE modeling system) was the only model studied by NOAA in the HIWPP study that did not use the shallow atmosphere approximation \citep{whitaker2015}. Using the deep atmosphere equations instead allows our simulations to include middle and upper atmospheric processes which are important for long-term (seasonal) weather and climate predictions. Furthermore, our code does not assume any special alignment of its mesh with the horizontal and vertical direction which allows the simulation of arbitrary steep terrain. It was possible to reach the desired resolution thanks to a careful optimization of the code and an excellent strong scaling efficiency of 99\% on the entire 3.14 million threads of the supercomputer Mira using a mesh with 1.8 billion grid points. To our knowledge, this paper not only presents the first atmospheric model that is capable of reaching the envisioned resolution within operational requirements, but also presents the first published strong scalability study up to petascale of fully compressible 3D global simulations.

\paragraph{Related Work}
\cite{johnsen2013petascale} present strong scaling efficiency of about 65\% at almost 300 TFlops sustained performance on the Cray machine Blue Waters for a Hurricane simulation using 4 billion grid points. \cite{wyszogrodzki2012parallel} present strong scaling up to $10^5$ cores on the Hopper II system including full parameterizations for moisture using up to 84 million grid points. Strong scaling for the atmospheric model CAM-SE using a spectral element method similar to the one utilized in NUMA is presented by \cite{dennisEtAl2012}. CAM-SE is targeted at climate prediction. Dennis et al. report strong scaling up to 172{,}800 cores on the Cray system JaguarPF using 81 million grid points. Other publications either do not solve the fully compressible Navier-Stokes equations \citep{tufo1999terascale,xue2014enabling} or show strong scaling only at much smaller scale \citep{Nair2009}. None of these publications is targeted at enabling numerical weather prediction at petascale.

Our paper is organized as follows: the numerical methods are introduced in Section \ref{sec:num_methods}. Section \ref{sec:testcases} presents the two test cases considered for the studies of this paper, Section \ref{sec:p4est} describes the mesh generation with the p4est library and Section \ref{sec:mira} gives some important technical details about the supercomputer Mira that we use for this work. A theoretical performance model is presented in Section \ref{sec:perfModel}. The code optimizations are presented in Section \ref{sec:optimizations} and scalability results are shown in Section \ref{sec:strong_scaling_efficiency}.


\section{Numerical Methods}
\label{sec:num_methods}

NUMA solves the compressible Navier-Stokes equations which can be written as \citep[see e.g.][]{Muller2013371}
\begin{align}
\label{eq:fluxform}
\frac{\partial\vek{q}}{\partial t}+\vek{\nabla}\cdot\matr{F}\left(\vek{q}\right)=\vek{S}\left(\vek{q}\right),
\end{align}
where $t$ is the time, \mbox{$\vek{q}\equiv\vek{q}(t,x,y,z)=\left(\rho, \rho\,\vek{u}^{\text{T}}, \Theta\right)^{\text{T}}$} is a time dependent vector field containing the so called prognostic variables (air density $\rho$, 3D wind speed, $\vek{u}$, and potential temperature, $\theta$) and $x, y, z$ are the coordinates in the three space dimensions. The nonlinear operator $\matr{F}$ denotes the flux tensor and $\vek{S}$ is a source function.

In the following subsections we illustrate the main steps of the numerical solution of these equations using a spectral element method. In the last subsection of this section we describe two different numerical possibilities to organize the data of our simulation. The two methods are identified as \emph{CG storage} and \emph{DG storage} \citep{abdi2016efficient}.

\subsection{Spatial Discretization}
\begin{figure}
\centering
\hspace{-3mm}\resizebox{9cm}{!}{\includegraphics{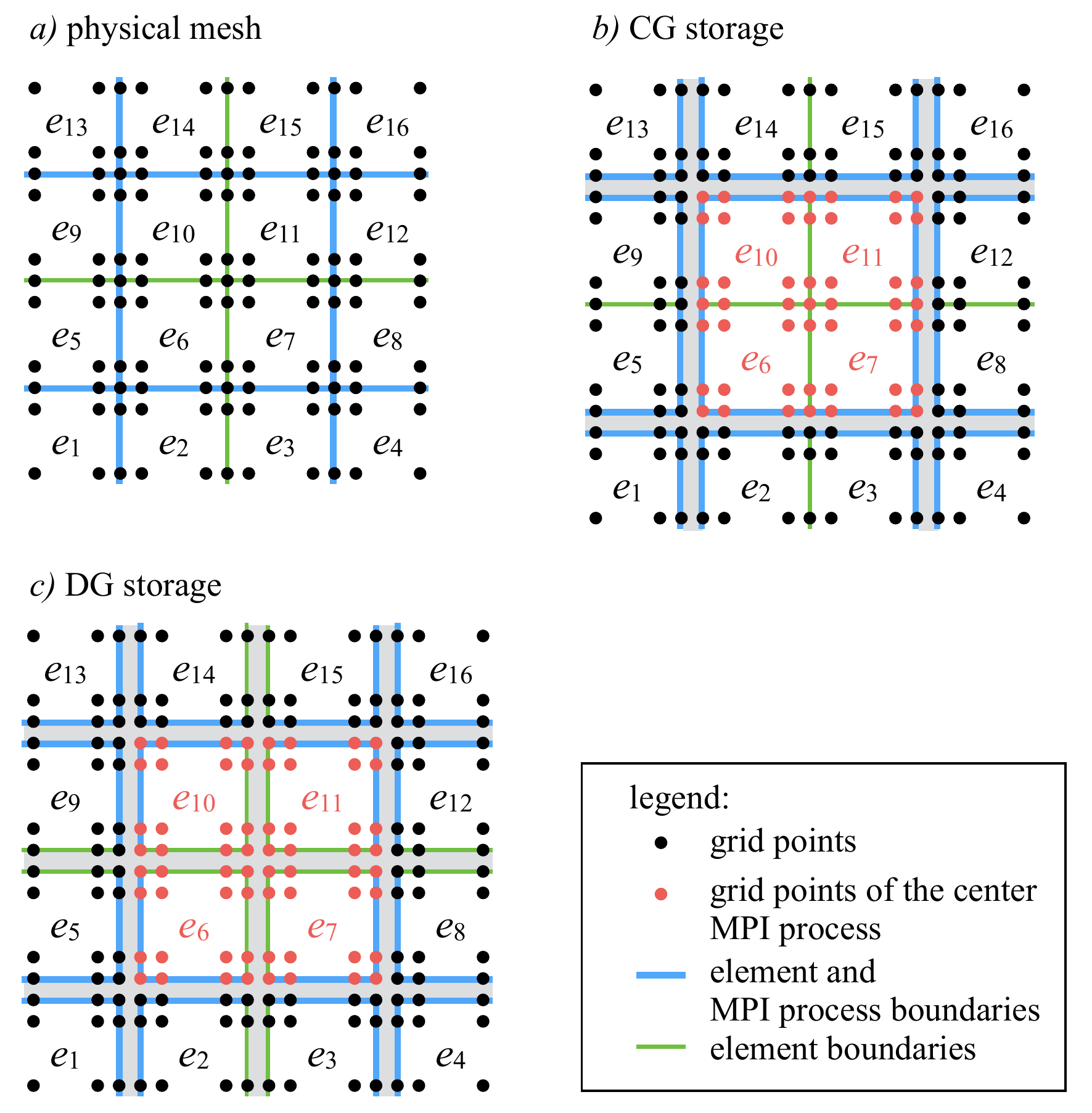}}
\caption{Illustration of a sample 2D cross section containing 16 elements ${e_1, \ldots, e_{16}}$ of our mesh (a) and two possible approaches to store the data: CG storage (b) and DG storage (c). For illustration purposes we assume that elements ${e_6}$, ${e_7}$, ${e_{10}}$ and ${e_{11}}$ are computed in the same MPI process while the other elements are computed on different MPI processes. The elements and grid points of the MPI process in the center of the figure are highlighted in red. The process boundaries are shown by the blue lines while element boundaries that are not process boundaries are shown by green lines. The square shape of the elements is used to keep this illustration simple. The elements can have arbitrarily curved faces.}
\label{fig:CGvsDGstorage}
\end{figure}

In order to discretize Eq. (\ref{eq:fluxform}) we introduce a mesh of elements. An example for a 2D cross section of our mesh is illustrated in Fig. \ref{fig:CGvsDGstorage}a. Inside each element we approximate the solution $\vek{q}$ in each dimension by polynomials of order $p$. We indicate with $\vek{q}_N$ the approximation of $\vek{q}$. To define these polynomials we introduce a mesh of $p+1$ grid points inside each element $e$ and in each direction. To simplify numerical integration we use Lobatto points. We denote the coordinates of these grid points by $(x_i, y_j, z_k)$. We restrict the rest of this section to the special case of $p=3$ because this case is most efficient for vectorization on the supercomputer Mira (see Section \ref{sec:optimizations}). We define our polynomials in a reference element over the interval $[-1,1]$ in each direction, with coordinates $(\xi,\eta,\zeta)$ and grid points $(\xi_i, \eta_j, \zeta_k)$. We denote the Jacobian determinant of the coordinate transformation between reference element and physical element $e$ at grid point $(x_i, y_j, z_k)$ with $J_{i,j,k}^e$. The numerical solution $\vek{q}_N$ inside element $e$ is given by
\begin{align}
\label{eq:expansion}
	\vek{q}_N(t,\xi,\eta,\zeta)=\sum_{l,m,n=1}^4 \vek{q}_{l,m,n}(t)\,\psi_n(\zeta)\psi_m(\eta)\psi_l(\xi),
\end{align}
with $\vek{q}_{l,m,n}(t)=\vek{q}(t,\xi_l,\eta_m,\zeta_n)$ and the 1D Lagrange basis polynomials $\psi_i$ are given by
\begin{align}
	\psi_i(\xi)=\begin{cases}\prod_{m\neq i}\frac{\xi-\xi_m}{\xi_i-\xi_m} &,~\xi\in[-1,1]\cr 0 &,~\xi\notin[-1,1],\end{cases}
\end{align}
where $\Omega_e$ is the domain of element $e$.

The goal is now to insert Eq. (\ref{eq:expansion}) into Eq. (\ref{eq:fluxform}) and solve it for the values of $\vek{q}_N$ at the grid points $\vek{q}_{i,j,k}$. In this paper we use a spectral element method. From now on, the spectral element method will be often referred to with the acronym CG, from Continuous Galerkin. We multiply Eq. (\ref{eq:fluxform}) by $\psi_i(\xi)\psi_j(\eta)\psi_k(\zeta)$ (including $J_{i,j,k}^e$)
and integrate over the entire domain $\Omega$. By using Gauss-Lobatto quadrature with quadrature weights $w_i$ we obtain the following equation:
\begin{align}
\label{eq:discstrongform}
M_{i,j,k}\frac{\D\vek{q}_{i,j,k}}{\dt}=-\sum_{e}J_{i,j,k}^e\,w_{i,j,k}\left(\vek{\nabla}\cdot\matr{F}_{N}-\vek{S}_{N}\right),
\end{align}
where \mbox{$w_{l,m,n}=w_l\,w_m\,w_n$}, \mbox{$\matr{F}_{N}=\matr{F}\left(\vek{q}_{N}\right)$}, \mbox{$\vek{S}_{N}=\vek{S}\left(\vek{q}_{N}\right)$} and $M_{i,j,k}=\sum_e w_{i,j,k}J_{i,j,k}^e$ are the entries of the diagonal mass matrix.

The spatial derivatives in the divergence of the flux tensor $\vek{\nabla}\cdot\matr{F}_{N}$ are given with Eq. (\ref{eq:expansion}) by:

\begin{align}
\label{eq:derivatives}
	\left.\frac{\partial \vek{q}_N}{\partial x}\right|_{i,j,k}&=\sum_{m=1}^4 \vek{q}_{m,j,k}\left.\frac{\D\psi_m(\xi)}{\D\xi}\right|_{\xi_i}\left.\frac{\partial\xi}{\partial x}\right|_{i,j,k}\nonumber\\&+\sum_{m=1}^4 \vek{q}_{i,m,k}\left.\frac{\D\psi_m(\eta)}{\D\eta}\right|_{\eta_i}\left.\frac{\partial\xi}{\partial x}\right|_{i,j,k}\\&+\sum_{m=1}^4 \vek{q}_{i,j,m}\left.\frac{\D\psi_m(\zeta)}{\D\zeta}\right|_{\zeta_i}\left.\frac{\partial\zeta}{\partial x}\right|_{i,j,k}.\nonumber
\end{align}
The products of the values of $\vek{q}_N$ at the grid points and the derivatives of our basis functions are essentially 4$\times$4 matrix-matrix-multiplications. All of the derivatives in Eq. (\ref{eq:derivatives}) are computed once at the beginning of the simulation.

The basis functions $\psi_i$ vanish outside of element $e$. For this reason the sum over all elements in Eq. (\ref{eq:discstrongform}) reduces to a single element for interior grid points. For the grid points along the edges we need to sum over all neighboring elements weighted by the volume of the elements (this summation is called DSS which stands for Direct Stiffness Summation). 

\subsection{Time Discretization}

In order to keep communication between different processors simple we use explicit time integration in the horizontal direction. If the vertical resolution is of the same order of magnitude of the horizontal resolution we use a fully explicit Runge-Kutta scheme with five stages and third order. In each of those five stages we need to evaluate the right hand side of Eq. (\ref{eq:discstrongform}) and communicate the values of the grid points along the process boundaries (blue lines in Fig. \ref{fig:CGvsDGstorage}a).

If the vertical resolution is much finer than the horizontal resolution we organize our mesh in such a way that all the vertical columns of our elements are always computed in the same MPI process. This allows us to make implicit corrections along the vertical columns after an explicit step of a leap-frog scheme. We call this approach 1D-IMEX (IMplicit-EXplicit) \citep{giraldo2013implicit}.

\subsection{Filter}

Spectral element methods require stabilization \citep{marras2015review}. NUMA allows for the use of different stabilization schemes that range from subgrid-scale models \citep{marras2015stabilized} to low-pass filters (Boyd-Vandeven). In this paper we use a Boyd-Vandeven filter. The main idea of this filter is to perform a spectral transformation of the nodal values $\vek{q}_{i,j,k}$ and to dampen the highest order modes. From a computational point of view this results in multiplying all the values $\vek{q}_{i,j,k}$ of the element $e$ with a filter matrix. Each time the filter is applied the new filtered values need to be communicated between neighboring MPI processes. In the future we will move to Laplacian based stabilization methods which do not require an additional communication step \citep[see][]{giraldo2013implicit}.

\subsection{CG and DG storage}

Each MPI process needs to own a copy of values at the grid points along the process boundaries. This is illustrated in Fig. \ref{fig:CGvsDGstorage}b by drawing a gray gap between the different processes. There is only one copy in memory for interior grid points even if they are located on a boundary between different elements (green lines in Fig. \ref{fig:CGvsDGstorage}b). We call this approach CG storage because it requires the solution to be continuous and works only for Continous Galerkin methods.

Another possibility to organize the data is to always store the values along element boundaries for each neighboring element separately (Fig. \ref{fig:CGvsDGstorage}c). We call this approach DG storage because it allows the use of Discontinous Galerkin methods. We compare the performance of these two approaches in a simple performance model in Section \ref{sec:perfModel}. 





\section{Test Cases}
\label{sec:testcases}

\begin{figure}
\centering
\hspace{-3mm}\resizebox{11cm}{!}{\includegraphics{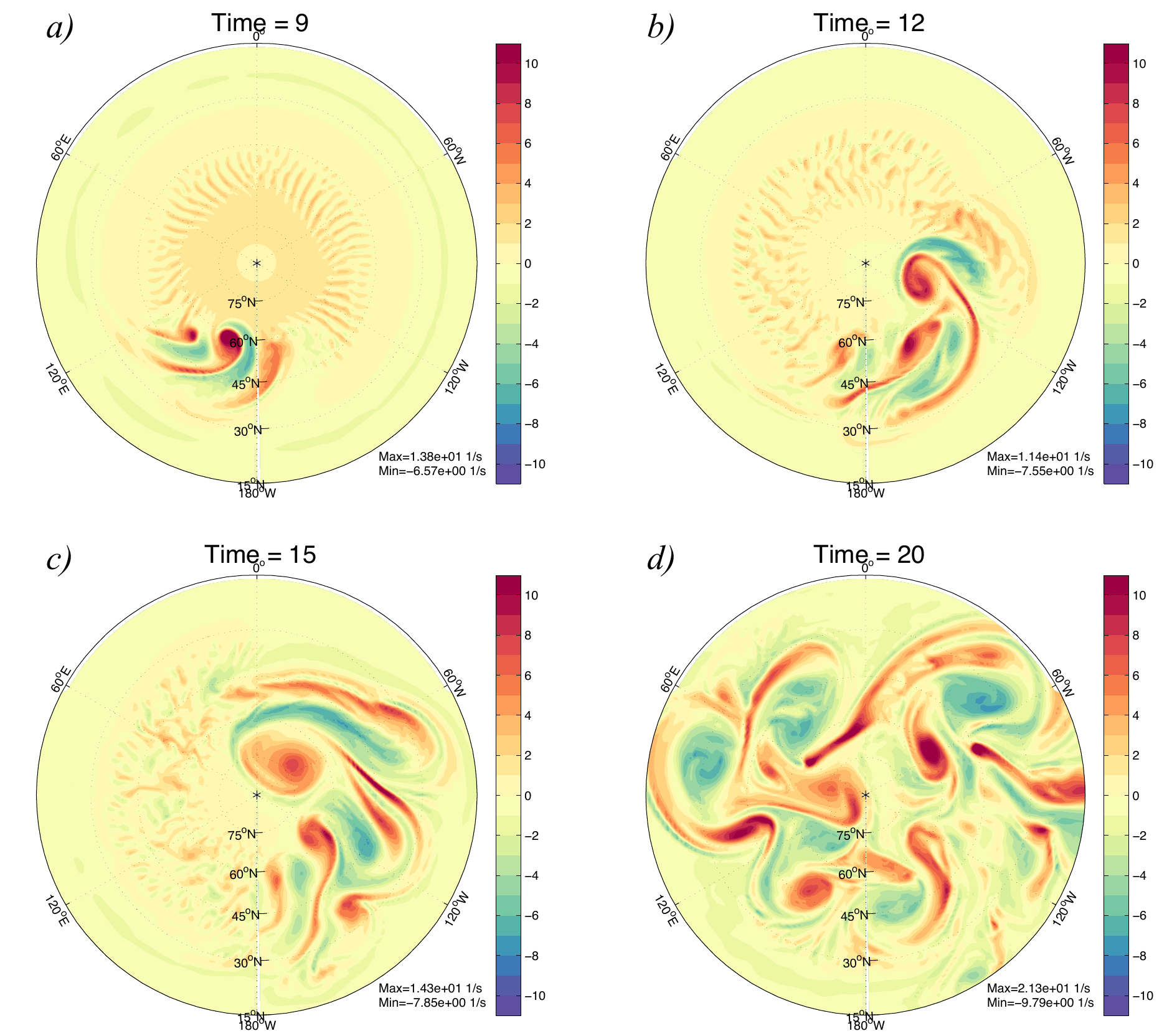}}
\caption{Baroclinic instability simulation at 9, 12, 15 and 20 days using a horizontal resolution of 50\,km and a vertical resolution of 1\,km. Shown is a top-view looking down onto the northern hemisphere. Plotted is the radial component of the vorticity $\vek{\omega}=\nabla\times\vek{u}$ in $\textrm{s}^{-1}$. This simulation uses a Large Eddy Simulation like stabilization method. We refer to \cite{marras2015review,marras2015stabilized} for more details about this stabilization method and to \cite{jablonowski2006baroclinic} for more details about the test case.}
\label{fig:baroclinicInstability}
\end{figure}

Two test cases are considered in this paper. One test case is the baroclinic wave instability problem by \cite{jablonowski2006baroclinic}. This problem is classically used to test the dynamical core of global circulation models \citep[also in HIWPP, cf.][]{hiwppProjectPlan}. It is initialized by a zonal band of high wind speed in the mid-latitudes (jet stream). A Gaussian perturbation of the zonal wind is added. This perturbation leads to wave like meridional perturbations of the jet stream (Fig. \ref{fig:baroclinicInstability}). After some time the flow pattern looks similar to the polar front jet stream of the real atmosphere.

The other test case is a 3D rising thermal bubble in a box of 1000$\,$m in each direction. This test case is initialized with a temperature perturbation in a neutrally stratified atmosphere. The precise definition and analysis of the full simulation is reported in \cite{Kelly2012}.

Both test cases are important for NWP applications. Operational weather prediction needs to cover the global circulation on the entire Earth like in the baroclinic wave test case. In order to use a higher resolution, for specific localized features of the atmosphere like hurricanes, one needs to run the simulation in limited area mode like in the 3D rising thermal bubble test case. 

\section{Mesh generation and load balancing}
\label{sec:p4est}

The data structures and algorithms for parallel mesh generation, partitioning, and load balancing used in our simulations were provided by the p4est library.
The p4est library has been used for efficient and scalable parallel adaptive mesh refinement for 2D advection on the sphere  \citep{BursteddeCalhounMandliEtAl14}, in other applications such as mantle convection and seismic wave propagation \citep{BursteddeGhattasGurnisEtAl10}, and as a backend for the deal.II finite element library \citep{BangerthHeisterHeltaiEtAl15}. Our present paper is the first time that p4est is used for full 3D atmospheric simulations.

The p4est library represents two- and three-dimensional domains via a two-level structure, with a macro mesh and a micro mesh.
The macro mesh is a conformal quadrilateral or hexahedral mesh, which is encoded as an unstructured mesh that is reproduced on each MPI process.
Each element in the macro mesh is then treated as the root of a partitioned quadtree or octree, which recursively refines the macro element isotropically to create a micro mesh.
The tree structure is represented in memory as a list of the leaves of the tree, ordered by the Morton curve (also known as the z-curve).
This ordering induces a space filling curve that visits the centers of the leaves: while this curve is not a continuous space filling curve, it has many of their nice properties.
One important property is that partitioning a domain by dividing the Morton curve into continuous segments creates subdomains that are fairly compact, with low surface-to-volume ratios \citep{HungershoferWierum02}.
This means that partitioning by this method keeps the intra-process communication during simulations low.
A full description of p4est's forest-of-quadtree and forest-of-octree data structures and algorithms can be found in \cite{BursteddeWilcoxGhattas11}.

When used in its raw form, the neighborhood information of an element in the micro mesh (i.e., which elements are adjacent) takes $\log(N_i)$ time to calculate, where $N_i$ is the number of micro mesh elements in the $i$th partition.
To avoid incurring this cost during each time step, the adjacency information for all elements in the $i$th partition can be converted into a lookup table, much like an unstructured mesh.
An efficient approach to creating this information, which can also be used to  enumerate the nodes for high-order CG finite elements, is described in \cite{IsaacBursteddeWilcoxEtAl15}.

The numerical methods in our simulations involve three-dimensional computations, but the radial direction is treated differently from the other two: its grid resolution requirements are different, and achieving efficiency in the IMEX time evolution scheme (and other calculations that are performed only in the radial direction) requires that radial columns of elements and degrees of freedom be contiguous in memory.
A forest-of-octrees approach would be ill-suited for these constraints.
First, octree refinement is isotropic: the aspect ratio of a macro element is inherited by all of the micro elements created by refinement. 
This means that the relationship between horizontal and radial resolution would have to be respected at the macro mesh level, increasing the macro mesh's complexity.
Second, the three-dimensional Morton curve does not respect the need to keep radial columns contiguous: elements in a column would be separated in memory, and without care would even be placed in separate partitions.

\begin{figure}\centering
  \resizebox{10cm}{!}{\includegraphics{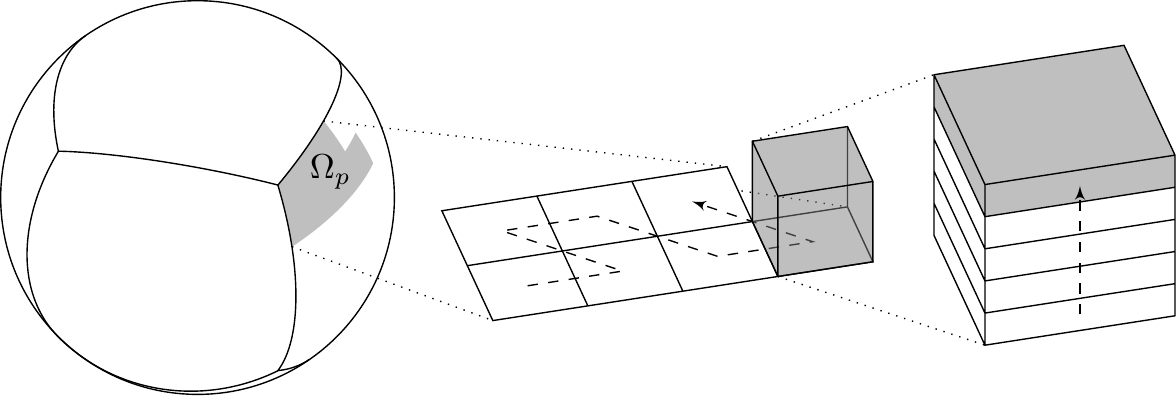}}
  \caption{%
    An illustration of meshing with the p6est extension of the p4est library.
    A macro mesh represents the cubed-sphere domain (left);
    division of the Morton curve creates the partitions for each MPI process;
    the columns in the $i$th partition $\Omega_i$ are ordered by a 2D Morton curve (middle);
    each radial column is stored as a list of ``layers'' from the bottom to the top (right).%
  }%
  \label{fig:p6est}
\end{figure}

For these reasons, we want to use a forest-of-quadtrees approach to generate and partition radial columns, but to handle the elements within each column using a different approach.
An extension to the p4est library, which was first used in the context of ice sheet modeling \citep{IsaacStadlerGhattas15}, provides the necessary data structures and interface. 
This extension is a set of ``p6est'' data types and functions (so named because it uses aspects of the two-dimensional ``p4est'' interface and the three-dimensional ``p8est'' interface).
Essentially, it treats each radial column as a list, from the bottom to top, of the elements created by recursive bisection of the full column, and uses the existing two-dimensional p4est routines to manage the partitioning of columns and intercolumn interactions.
A more complete description of this approach can be found in \cite[Chapter 2]{Isaac15}.
The p6est mesh format is illustrated in Figure~\ref{fig:p6est}.

\begin{figure}\centering
\resizebox{9cm}{!}{\includegraphics{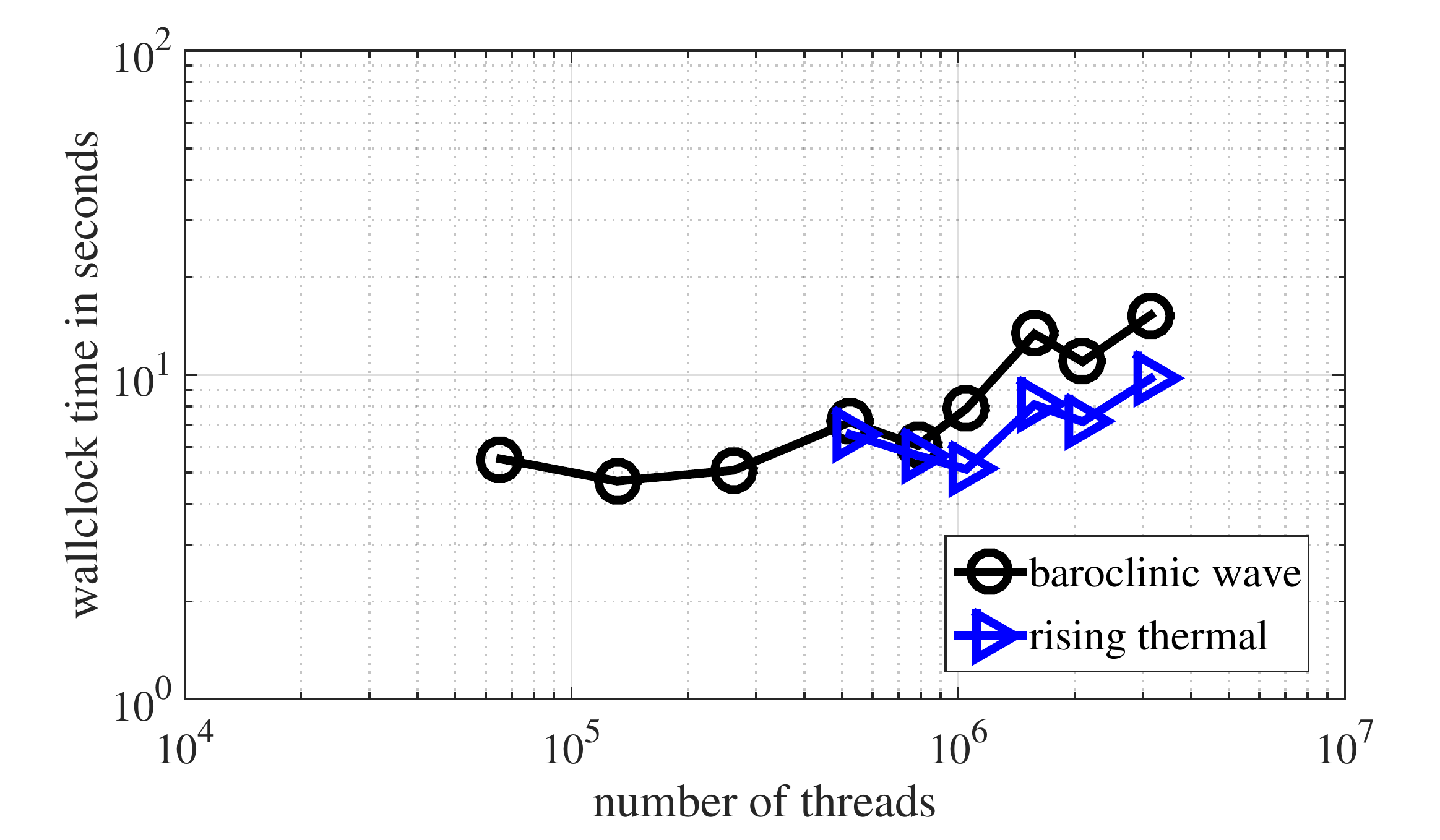}}
\caption{Wallclock time in seconds of the mesh generation with the p4est library up to the entire machine Mira for the simulations in Fig. \ref{fig:modeldays} and \ref{fig:rtbScaling}.}
\label{fig:p4est}
\end{figure}

As the elements within a column are defined by recursive bisection, p6est was designed for meshes with $2^k$ elements per column for some $k$.
Because NUMA must work with meshes which do not have this property, the p6est format was extended for this work to support an arbitrary number of elements per column. The runtime of the mesh generation is shown in Fig. \ref{fig:p4est}. Even for 43 billion grid points (blue results) it takes always less than 20 seconds runtime to generate the mesh. 

It should be noted that, although mesh adaptivity is not used in this work, the p6est data structures support bimodal local mesh adaptivity: elements may be independently refined in the radial direction, and each column can be independently refined into four smaller columns.

\section{Blue Gene Q Mira}
\label{sec:mira}

The simulations presented in this paper were performed on the supercomputer Mira of the Argonne National Laboratory. Mira is an IBM Blue Gene/Q system offering $49{,}152$ computational nodes. Each of these nodes has 16 cores resulting in a total number of $786{,}432$ cores. Each core has a quad floating point unit. This permits running up to four MPI processes or up to four OpenMP threads on each core. The maximum total number of hardware threads is therefore $3{,}145{,}728$.

Important for the performance of our code is the memory architecture of Mira. Each computational node has 16$\,$GB of random access memory (RAM). The processor receives its data from RAM through two levels of cache. Each core has its own level 1 (L1) cache of 16$\,$KB while the L2 cache of Mira is shared among all 16 processors of the computational node and has a size of 32$\,$MB.

In addition to these two levels of cache, each core has a L1 cache prefetch (L1P) of 4KB \citep{chung2012application}. Whenever the data needed for the simulation cannot be found in the L1 cache (L1 cache miss) the computer checks if this data is available in L1P. This is always done for an entire cache line of 128 bytes (16 double precision floating point numbers). If the requested cache line is not in L1P it goes on and looks in L2 cache and eventually in the RAM for this data. The prefetcher of the Blue Gene/Q keeps track of previous cache misses. The stream prefetcher of the BG/Q establishes a stream if consecutive cache lines are requested, i.e. L1P loads the next cache lines even when no cache miss occurs. If this consecutive data is actually needed by the code this can save a lot of runtime. However, if the data is not used consecutively by the code the prefetcher will read a lot of data from L2 cache without ever using it. This can produce a huge number of unnecessary cache misses.

Important for our optimizations is also the vector unit. The registers of Mira have a length of 256 bits offering space for four double precision floating point numbers. This allows performing up to 8 double precision fused multiply-add floating point operations per core within the same clock cycle. Each core can perform one floating point instruction and one integer instruction per clock cycle. Load and store instructions take each one cycle of the integer unit.

\section{Performance Model}
\label{sec:perfModel}

Making a theoretical performance model allows us to estimate the expected performance and to compare different numerical methods without fully optimizing them in our code. We created the performance model by counting all floating point operations as well as memory read and write accesses throughout our entire code. The results of this theoretical model are presented in Tables \ref{tab:perfModelCase21NoRandAcc} to \ref{tab:perfModelCase61}. We do this for three possible implementations: our CG version avoids having multiple copies of data in memory wherever possible (as in Fig.\ref{fig:CGvsDGstorage}b). The only variables that still require multiple values at the same physical locations are the metric terms in Eq. (\ref{eq:derivatives}). The CG/DG-version uses the same approach like the CG version for dynamically changing variables but allows additional copies of the data at element boundaries (Fig. \ref{fig:CGvsDGstorage}c) for data related to the reference atmosphere that does not change throughout the simulation. The DG-version allows multiple copies of the data along element interfaces (Fig. \ref{fig:CGvsDGstorage}c) for all variables.

Table \ref{tab:perfModelCase21NoRandAcc} shows the expected number of floating point operations and total memory read and write traffic throughout the entire simulation for the rising thermal bubble test case. The additional copies of data along the element interfaces in DG storage lead to a significant increase in the number of floating point operations as well as memory traffic and to a significant reduction of arithmetic intensity. Our results for arithmetic intensity show that all of our cases are memory bound (see roofline plot in Fig. \ref{fig:roofline21} and \ref{fig:roofline61}). The optimal runtime in Table \ref{tab:perfModelCase21NoRandAcc} is the runtime that we expect if we manage to reach 100\% of the theoretical peak memory bandwidth as given by the STREAM benchmark in \cite{morozov2013early}. The percentage of the theoretical peak performance of the processor is also given in Table \ref{tab:perfModelCase21NoRandAcc} for this optimal runtime. Our CG-only version gives us the best performance in all of these results.

\begin{table*}\centering
\scriptsize
\begin{tabular}{!{\color{black}\vrule}l!{\color{black}\vrule}c!{\color{black}\vrule}c!{\color{black}\vrule}c!{\color{black}\vrule}}
	\hline
	& \textbf{CG} & \textbf{CG/DG} & \textbf{DG}\\
	\hline
	GFlops per node & 3007.00 & 3007.00 & 4023.19\\
	read traffic in GB & 2129.42 & 2537.28 & 3489.66\\
	write traffic in GB & 661.83 & 688.34 & 1168.69\\
	arithmetic intensity in Flops/Bytes & 1.08 & 0.93 & 0.86\\
	optimal runtime in seconds & 97.94 & 113.18 & 163.45\\
	\% of theoretical peak of processor & 14.99 & 12.97 & 12.02\\
	\arrayrulecolor{black}
	\hline
\end{tabular}
\vskip 0.2cm
\caption{Theoretical amount of floating point operations and memory traffic for the rising thermal bubble test case at polynomial degree p=3. The effect of random memory access for CG is not taken into account in these results. All the numbers given in this table are summed over the entire simulation on 768 BG/Q nodes. The simulations use an effective resolution of $1.30\,\text{m}$ in x and y direction and an effective resolution of $0.89\,\text{m}$ in z direction which leads to a total number of 7.4${\times10^{8}}$ grid points. This corresponds to 60k grid points per BG/Q core which is the workload we aim at using on the entire machine Mira. All simulations use a Courant number of 0.7 in the vertical direction and are run for a model time of 1 second. This corresponds for p=3 to 690 timesteps. The column ``CG'' and ``DG'' give the results for CG and DG storage respectively. The column ``CG/DG'' uses DG storage for data that does not change throughout the simulation and CG storage for the rest. The number of floating point operations is adjusted according to real measurements to take the effects of compiler optimization into account. The optimal runtime is based on the peak memory bandwidth of 28.5 GB/s according to the STREAM benchmark \citep{morozov2013early}. This runtime leads to the given percentage of the theoretical peak performance of the processor. This demonstrates that all of our simulations are expected to be memory bound.}
\label{tab:perfModelCase21NoRandAcc}
\end{table*}

CG-storage has a big disadvantage that we have not included in Table \ref{tab:perfModelCase21NoRandAcc}: we have to perform the computations of Eq. (\ref{eq:derivatives}) on a per element basis while data stored in CG-storage is not arranged on a per element basis. This leads to non-contiguous memory access that appears to be random. Having random memory access will not affect the results in Table \ref{tab:perfModelCase21NoRandAcc} if most of the data is already in L2 cache. Therefore the results of Table \ref{tab:perfModelCase21NoRandAcc} will still be correct if the number of elements per compute core is small. Our goal for the rising thermal bubble problem is to use a very high grid resolution which produces much more data than we can fit into L2 cache. In this case we expect to get a much better estimate for the performance of our code by assuming that none of the data is already in L2 cache and by counting the full cache lines in those cases where CG storage requires us to access only a small portion of that cache line. The results including this estimate for the effect of random memory access are shown in Table \ref{tab:perfModelCase21WithRandAcc}. Including the effect of random memory access does not change the number of floating point operations but it leads to a significant increase in memory traffic. This makes DG storage now the version with the highest arithmetic intensity and therefore the best percentage of the theoretical peak performance of the processes. The overall runtime of the entire simulation is still slower for DG storage due to the increased number of floating point operations. The winner in terms of overall runtime is our CG/DG-version.

\begin{table*}\centering
\scriptsize
\begin{tabular}{!{\color{black}\vrule}l!{\color{black}\vrule}c!{\color{black}\vrule}c!{\color{black}\vrule}c!{\color{black}\vrule}}
	\hline
	& \textbf{CG} & \textbf{CG/DG} & \textbf{DG}\\
	\hline
	GFlops per node & 3007.00 & 3007.00 & 4023.19\\
	read traffic in GB & 3483.05 & 3138.46 & 3682.77\\
	write traffic in GB & 853.44 & 879.95 & 1360.30\\
	arithmetic intensity in Flops/Bytes & 0.69 & 0.75 & 0.80\\
	optimal runtime in seconds & 152.16 & 141.00 & 176.95\\
	\% of theoretical peak of processor & 9.65 & 10.41 & 11.10\\
	\arrayrulecolor{black}
	\hline
\end{tabular}
\vskip 0.2cm
\caption{Like Table \ref{tab:perfModelCase21NoRandAcc} but including an estimate for the effect of random memory access due to CG storage if the previously computed elements are no longer available in cache.}
\label{tab:perfModelCase21WithRandAcc}
\end{table*}

\begin{figure}\centering
\resizebox{\textwidth}{!}{\includegraphics{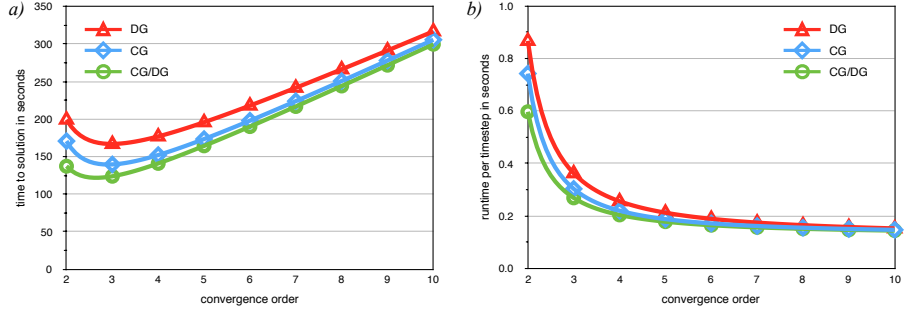}}
\caption{Time to solution (a) and runtime per timestep (b) as a function of convergence order $p+1$ for the CG and DG storage versions and setup shown in Table \ref{tab:perfModelCase21WithRandAcc} for the rising thermal bubble test case. All polynomial orders use the same effective resolution (average distance between points) of $1.3\,$m in the horizontal direction and $0.89\,$m in the vertical direction and the same Courant number.}
\label{fig:perfModel}
\end{figure}

All of the results presented so far have been obtained for $p=3$. Our performance model allows us to compare the optimal performance of different polynomial orders $p$ for the three different versions from Table \ref{tab:perfModelCase21WithRandAcc}. The results are shown in Fig. \ref{fig:perfModel}. The runtime per timestep increases with decreasing order due to the additional copies of data for DG storage (Fig. \ref{fig:perfModel}b). This effect occurs also in our CG-version because even in this version we still need to store metric terms in DG-storage.
The results look very different when the actual time to solution for the entire simulation is considered (Fig. \ref{fig:perfModel}a). The numerical methods considered in this paper require to reduce the time step with increasing order to keep the explicitly resolved part of the simulation stable. This makes low polynomial order much faster than high order. As shown in Fig. \ref{fig:perfModel}a we get the most efficient time to solution for convergence order 3 which corresponds to polynomial order $p=2$. We still choose $p=3$ for the remainder of this paper because we expect $p=3$ to be easier to vectorize on the Blue Gene Q architecture.

Table \ref{tab:perfModelCase61} shows the results of our theoretical performance model for the baroclinic instability test case. As described in the introduction our final goal is to run this test case at 3km resolution on the entire supercomputer Mira. This setup gives us only 20 elements per hardware thread which makes it possible to fit most of the data in each timestep into L2 cache. For this reason we do not estimate the effect of random memory access in this case. Our CG and CG/DG versions are again significantly faster than our DG version. We will use the CG/DG version for our optimizations in the next section because this version gives us a significant advantage for the rising thermal bubble test case in Table \ref{tab:perfModelCase21WithRandAcc}.

\begin{table*}\centering
\scriptsize
\begin{tabular}{!{\color{black}\vrule}l!{\color{black}\vrule}c!{\color{black}\vrule}c!{\color{black}\vrule}c!{\color{black}\vrule}}
	\hline
	& \textbf{CG} & \textbf{CG/DG} & \textbf{DG}\\
	\hline
	GFlops per node & 61.96 & 61.96 & 83.00\\
	read traffic in GB & 58.55 & 62.18 & 94.22\\
	write traffic in GB & 22.48 & 22.48 & 36.69\\
	arithmetic intensity in Flops/Bytes & 0.76 & 0.73 & 0.63\\
	optimal runtime in seconds & 2.84 & 2.97 & 4.59\\
	\% of theoretical peak of processor & 10.64 & 10.18 & 8.82\\
	\arrayrulecolor{black}
	\hline
\end{tabular}
\vskip 0.2cm
\caption{Like Table \ref{tab:perfModelCase21NoRandAcc} but for the baroclinic instability test case. These results are calculated for 972 BG/Q nodes. The simulations use an effective horizontal resolution of $21.3\,\text{km}$ and an effective vertical resolution of $1.0\,\text{km}$ which leads to a total number of 4.4${\times10^{7}}$ grid points. This corresponds to 2821 grid points per BG/Q core which is the workload we aim at using on the entire machine Mira. All simulations use a Courant number of 0.4 in the horizontal direction and 6.4 in the vertical direction and are run for a model time of 4 hours. This corresponds for p=3 to 947 timesteps.}
\label{tab:perfModelCase61}
\end{table*}

\section{Code Optimizations}
\label{sec:optimizations}

The goal of this section is to optimize our CG storage version of NUMA for $p=3$. To take advantage of the reduced amount of data compared to DG storage we aim at computing as much work as possible on a per grid point basis and try to avoid making computations on a per element basis. The main structure of our code is illustrated in code example \ref{lst:struct}. Computations that need to be computed element-wise are highlighted in blue. Communication is highlighted in red.

\begin{code}
\begin{algorithm}
 create the mesh
 initialize data q at time t=0
 while time < final time
   begin loop over time integration stages
     @\color{blue}compute right hand side of Eq. (\ref{eq:discstrongform}) for current stage ({\textit{create\_rhs}})@
     @\color{red}{{communicate}}@
     DSS @and@ multiply with inverse mass matrix
     update q
   end
   compute @{IMEX}@ corrections for each vertical column of grid points
   @\color{blue}compute {filter}@
   @\color{red}{{communicate}}@
   DSS @and@ multiply with inverse mass matrix
   increment time
 end
\end{algorithm}
\caption{Pseudocode of the main structure of our code NUMA. The part of the code highlighted in blue needs to be computed element-wise. The rest (black lines) can be computed for each grid point separately. MPI communication is highlighted in red.}
\label{lst:struct}
\end{code}

We tried to optimize all parts of our code. We found \textit{create\_rhs} (the computation of the right hand side in Eq.(\ref{eq:discstrongform}), see also code example \ref{lst:struct}) to be the only function that contains enough floating point operations to allow significant optimizations.

Tables \ref{tab:optimizations21:createrhs1} to \ref{tab:optimizations61:timeloop1} show performance measurements for different versions of \textit{create\_rhs}. The different versions are explained in Table \ref{tab:optimizationVersions}. For the rest of this section we simply refer to the different versions in these tables. Version T gives theoretical expectations from our performance model as presented in the previous section. As in Table \ref{tab:perfModelCase21WithRandAcc} we include here our estimate for random memory access for the rising thermal bubble test case. Version O is another theoretical result that is obtained by making some further optimizations. These optimizations consist of avoiding some unnecessary memory access and minimizing memory access by computing metric terms in Eq. (\ref{eq:derivatives}) in each stage of our time integration method. This leads to a significant increase in the number of floating point operations but more importantly it allows a significant reduction in memory traffic. We will try these optimizations in our future work.

All of the measurements in Tables \ref{tab:optimizations21:createrhs1} to \ref{tab:optimizations61:timeloop1} are taken over the entire simulation. This includes that $p=6$ (version A) requires a larger number of timesteps than $p=3$ (version B). The column ``\% peak'' gives the percentage of the theoretical peak performance of the processor. The column ``arith. int.'' shows the arithmetic intensity in Flops/Bytes. The column ``\% max.'' shows how close this part of the code is to the maximum performance according to the roofline model for the given arithmetic intensity. The three columns ``flops'', ``read'' and ``write'' show the total number of flops, read and write traffic summed over the entire simulation on one BG/Q node. The column ``vect.'' shows how many percent of all floating point operations are vectorized. The column ``fma'' gives the percentage of fused multiply-add operations among all floating point operations. We have not estimated the optimal number of fused multiply-add operations in our performance model. For this reason we leave the column ``fma'' empty for the versions T and O. The column ``mix'' shows the percentage of floating point instructions among all instructions. The column ``issue'' shows how close this part of the code is to the maximum issue rate of one integer/load/store instruction per cycle per core. The three columns ``L1P'', ``L2'' and ``DDR'' show how many of the loads hit the L1P buffer, the L2 cache and the DDR memory respectively.

To confirm our theoretical results from Fig. \ref{fig:perfModel} we first tried a fairly high polynomial order of $p=6$ (version A). In agreement with our theoretical results we found that order $p=3$ gives us significantly better time to solution (version B). We have not seen a significant impact on the accuracy of our test cases. We use $p=3$ for the rest of this paper because this order is very well suited for vectorization on Mira (four double precision floating point numbers fit into one register). Another significant speedup was obtained by running four MPI processes per core (version C).

We computed the derivatives of Eq. (\ref{eq:derivatives}) in versions A, B and C for each of the five variables of $\vek{q}$ separately. By merging all these derivatives into one matrix for each direction and element we achieved another significant speedup (version D). We also used the BLAS function dgemm (version E) but without any improvement on the performance.

The rest of our optimizations can be categorized into three main topics which we discuss in the following subsections: compiler optimizations, BG/Q vector intrinsics and OpenMP. At the end of this section we give a short description of possible next steps for further optimization.

\subsection{Compiler Optimization}

To improve the performance while retaining portability we worked first on enabling better optimization through the compiler. We spent some time on finding the best level of compiler optimization for each function of our code. We found a few functions for which level 3 optimization gave us wrong results. This is not surprising because level 3 compiler optimization is not IEEE compliant.

Many of our operations in \textit{create\_rhs} looked initially like code example \ref{lst:noopt}. The operations were computed for each grid point of the element separately which makes it impossible for the compiler to vectorize the code. This explains the very low fraction of vectorized operations in versions A, B, C and D (column ``vect.'' in Table \ref{tab:optimizations21:createrhs1} to \ref{tab:optimizations61:timeloop1}).
\begin{table*}\centering
\scriptsize
\begin{tabular}{!{\color{black}\vrule}c!{\color{black}\vrule}l!{\color{black}\vrule}}
	\hline
	\textbf{version} & \textbf{description}\\
	\hline
A & $p=6$, 1 MPI process per core\\
B & $p=3$, 1 MPI process per core\\
C & like B, 4 MPI processes per core\\
D & like C, all derivatives of $e$ in one matrix\\
E & like D, use BLAS function dgemm\\
F & like C, optimized for compiler vectorization\\
\hline
G & like D, some vector intrinsics\\
H & like C, rewritten using vector intrinsics\\
I & like H, 4 OpenMP threads per core\\
\hline
T & theoretical expectation for version I\\
O & like T, further optimized (see text for details)\\
	\arrayrulecolor{black}
	\hline
\end{tabular}
\vskip 0.2cm
\caption{Description of the different optimization stages shown in Tables \ref{tab:optimizations21:createrhs1} to \ref{tab:optimizations61:timeloop1}. Versions A to F are platform independent whereas versions G to I make use of special BG/Q vector instructions. The versions T and O are theoretical results based on counting memory access by hand and estimating the runtime by using the memory bandwidth according to the STREAM benchmark results in \cite{morozov2013early}.}
\label{tab:optimizationVersions}
\end{table*}
\begin{landscape}
\begin{table*}\centering
\scriptsize
\begin{tabular}{c!{\color{black}\vrule}c!{\color{black}\vrule}c!{\color{black}\vrule}c!{\color{black}\vrule}c!{\color{black}\vrule}c!{\color{black}\vrule}cc!{\color{black}\vrule}cc!{\color{black}\vrule}cc!{\color{black}\vrule}ccc}
& & & \bfseries arith. & & & \multicolumn{2}{c|}{\bfseries traffic in GB} & \multicolumn{2}{c|}{\bfseries flops} & \multicolumn{2}{c|}{\bfseries instructions} & \multicolumn{3}{c}{\bfseries loads that hit} \\

& \bfseries runtime & \bfseries \% peak & \bfseries int. & \bfseries \% max. & \bfseries GFlops & \bfseries read & \bfseries write & \bfseries vect. & \bfseries fma & \bfseries mix & \bfseries issue & \bfseries L1P & \bfseries L2 & \bfseries DDR \\
\hline
\bfseries A & 855.1$\,$s & 2.0$\,$\% & 1.2 & 12.0$\,$\% & 3419.4 & 2466.4 & 469.8 & 31.1$\,$\% & 85.1$\,$\% & 23.0$\,$\% & 39.5$\,$\% & 3.7$\,$\% & 1.8$\,$\% & 0.2$\,$\%\\
\bfseries B & 802.2$\,$s & 1.4$\,$\% & 1.1 & 8.8$\,$\% & 2315.3 & 1753.3 & 293.1 & 14.4$\,$\% & 80.0$\,$\% & 18.0$\,$\% & 36.2$\,$\% & 1.4$\,$\% & 3.6$\,$\% & 0.2$\,$\%\\
\bfseries C & 351.4$\,$s & 3.2$\,$\% & 0.9 & 25.6$\,$\% & 2315.8 & 2111.4 & 481.5 & 14.4$\,$\% & 80.0$\,$\% & 18.0$\,$\% & 67.7$\,$\% & 2.6$\,$\% & 7.0$\,$\% & 0.5$\,$\%\\
\bfseries D & 188.8$\,$s & 5.9$\,$\% & 1.0 & 44.2$\,$\% & 2280.1 & 1910.3 & 513.7 & 33.0$\,$\% & 79.4$\,$\% & 26.5$\,$\% & 61.1$\,$\% & 1.8$\,$\% & 8.0$\,$\% & 0.6$\,$\%\\
\bfseries E & 337.8$\,$s & 3.5$\,$\% & 0.9 & 29.1$\,$\% & 2412.7 & 2197.9 & 703.5 & 51.0$\,$\% & 85.8$\,$\% & 16.9$\,$\% & 53.6$\,$\% & 2.2$\,$\% & 11.2$\,$\% & 0.6$\,$\%\\
\bfseries F & 180.4$\,$s & 6.6$\,$\% & 1.0 & 48.8$\,$\% & 2438.6 & 2032.0 & 521.0 & 73.9$\,$\% & 81.3$\,$\% & 21.2$\,$\% & 49.3$\,$\% & 4.6$\,$\% & 14.4$\,$\% & 0.5$\,$\%\\
\hline
\bfseries G & 102.1$\,$s & 12.0$\,$\% & 1.1 & 74.8$\,$\% & 2503.8 & 1776.5 & 493.7 & 86.4$\,$\% & 75.7$\,$\% & 27.0$\,$\% & 50.6$\,$\% & 7.5$\,$\% & 6.6$\,$\% & 0.7$\,$\%\\
\bfseries H & 88.1$\,$s & 13.9$\,$\% & 1.2 & 81.6$\,$\% & 2503.7 & 1677.5 & 488.3 & 98.6$\,$\% & 75.7$\,$\% & 28.9$\,$\% & 39.7$\,$\% & 7.2$\,$\% & 12.3$\,$\% & 0.8$\,$\%\\
\bfseries I & 96.7$\,$s & 12.6$\,$\% & 1.3 & 71.3$\,$\% & 2503.4 & 1657.6 & 382.5 & 98.6$\,$\% & 75.7$\,$\% & 28.1$\,$\% & 39.9$\,$\% & 7.6$\,$\% & 13.9$\,$\% & 0.7$\,$\%\\
\hline
\bfseries T & 86.2$\,$s & 14.2$\,$\% & 1.0 & 100.0$\,$\% & 2503.4 & 2165.5 & 291.7 & 100.0$\,$\% & & 50.0$\,$\% & 100.0$\,$\% & 0.0$\,$\% & 0.0$\,$\% & 0.0$\,$\% \\
\bfseries O & 41.5$\,$s & 36.7$\,$\% & 2.6 & 100.0$\,$\% & 3119.1 & 1051.3 & 132.5 & 100.0$\,$\% & & 50.0$\,$\% & 100.0$\,$\% & 0.0$\,$\% & 0.0$\,$\% & 0.0$\,$\% \\
	\arrayrulecolor{black}
\end{tabular}
\vskip 0.2cm
\caption{Performance measurements for \textit{create\_rhs} (the computation of the right hand side in Eq.(\ref{eq:discstrongform}), see also code example \ref{lst:struct}) with the Hardware Performance Monitor Toolkit for the rising thermal bubble test case and the different versions of our code as described in Table \ref{tab:optimizationVersions}. We use the setup as described in Table \ref{tab:perfModelCase21NoRandAcc}. Version A to I are measured by using the Hardware Performance Monitoring Toolkit whereas versions T and O are theoretical results based on counting memory access by hand. Version T includes estimates for the effect of random memory access due to non-optimal partitioning of our mesh. The column ``runtime'' shows the overall time spent in \textit{create\_rhs} (over all 690 timesteps for p=3 (3450 executions of \textit{create\_rhs}) and over 1123 timesteps for p=6 (5615 executions of \textit{create\_rhs})). All measurements were obtained by taking the average values over four arbitrary nodes. We refer to the main text for the explanation of the other columns.}
\label{tab:optimizations21:createrhs1}
\end{table*}
%
\begin{table*}\centering
\scriptsize
\begin{tabular}{c!{\color{black}\vrule}c!{\color{black}\vrule}c!{\color{black}\vrule}c!{\color{black}\vrule}c!{\color{black}\vrule}c!{\color{black}\vrule}cc!{\color{black}\vrule}cc!{\color{black}\vrule}cc!{\color{black}\vrule}ccc}
& & & \bfseries arith. & & & \multicolumn{2}{c|}{\bfseries traffic in GB} & \multicolumn{2}{c|}{\bfseries flops} & \multicolumn{2}{c|}{\bfseries instructions} & \multicolumn{3}{c}{\bfseries loads that hit} \\

& \bfseries runtime & \bfseries \% peak & \bfseries int. & \bfseries \% max. & \bfseries GFlops & \bfseries read & \bfseries write & \bfseries vect. & \bfseries fma & \bfseries mix & \bfseries issue & \bfseries L1P & \bfseries L2 & \bfseries DDR \\
\hline
\bfseries A & 1099.9$\,$s & 1.9$\,$\% & 0.7 & 20.1$\,$\% & 4285.1 & 4501.5 & 1812.0 & 28.9$\,$\% & 83.7$\,$\% & 22.8$\,$\% & 38.9$\,$\% & 4.0$\,$\% & 1.3$\,$\% & 0.4$\,$\%\\
\bfseries B & 972.7$\,$s & 1.4$\,$\% & 0.7 & 14.8$\,$\% & 2810.4 & 2995.7 & 1100.9 & 13.6$\,$\% & 79.1$\,$\% & 17.9$\,$\% & 36.8$\,$\% & 1.9$\,$\% & 2.9$\,$\% & 0.4$\,$\%\\
\bfseries C & 454.3$\,$s & 3.0$\,$\% & 0.6 & 34.9$\,$\% & 2833.3 & 3299.9 & 1220.2 & 13.6$\,$\% & 79.1$\,$\% & 17.1$\,$\% & 69.1$\,$\% & 2.8$\,$\% & 6.0$\,$\% & 0.6$\,$\%\\
\bfseries D & 293.0$\,$s & 4.7$\,$\% & 0.6 & 52.5$\,$\% & 2796.7 & 3167.8 & 1221.4 & 29.0$\,$\% & 78.6$\,$\% & 22.3$\,$\% & 65.2$\,$\% & 2.4$\,$\% & 6.4$\,$\% & 0.8$\,$\%\\
\bfseries E & 447.8$\,$s & 3.2$\,$\% & 0.7 & 34.1$\,$\% & 2928.2 & 3131.2 & 1218.8 & 44.1$\,$\% & 83.9$\,$\% & 15.6$\,$\% & 60.8$\,$\% & 2.5$\,$\% & 9.0$\,$\% & 0.6$\,$\%\\
\bfseries F & 283.6$\,$s & 5.1$\,$\% & 0.7 & 56.2$\,$\% & 2956.1 & 3318.2 & 1220.4 & 63.1$\,$\% & 80.2$\,$\% & 18.4$\,$\% & 57.4$\,$\% & 4.4$\,$\% & 9.9$\,$\% & 0.7$\,$\%\\
\hline
\bfseries G & 205.0$\,$s & 7.2$\,$\% & 0.7 & 77.3$\,$\% & 3020.8 & 3296.9 & 1221.2 & 73.8$\,$\% & 75.6$\,$\% & 20.8$\,$\% & 60.3$\,$\% & 5.9$\,$\% & 4.5$\,$\% & 0.8$\,$\%\\
\bfseries H & 191.7$\,$s & 7.7$\,$\% & 0.7 & 82.5$\,$\% & 3020.4 & 3285.3 & 1221.8 & 84.0$\,$\% & 75.6$\,$\% & 20.5$\,$\% & 55.5$\,$\% & 5.6$\,$\% & 7.0$\,$\% & 1.0$\,$\%\\
\bfseries I & 198.5$\,$s & 7.4$\,$\% & 0.7 & 72.7$\,$\% & 3007.5 & 3009.6 & 1102.1 & 83.5$\,$\% & 75.0$\,$\% & 21.6$\,$\% & 48.7$\,$\% & 6.0$\,$\% & 7.3$\,$\% & 1.0$\,$\%\\
\hline
\bfseries T & 141.0$\,$s & 10.4$\,$\% & 0.7 & 100.0$\,$\% & 3007.5 & 3138.5 & 879.9 & 100.0$\,$\% & & 50.0$\,$\% & 100.0$\,$\% & 0.0$\,$\% & 0.0$\,$\% & 0.0$\,$\%  \\
\bfseries O & 80.8$\,$s & 21.9$\,$\% & 1.6 & 100.0$\,$\% & 3623.1 & 1799.4 & 502.8 & 100.0$\,$\% & & 50.0$\,$\% & 100.0$\,$\% & 0.0$\,$\% & 0.0$\,$\% & 0.0$\,$\% \\
	\arrayrulecolor{black}
\end{tabular}
\vskip 0.2cm
\caption{Performance measurements for the rising thermal bubble test case like in Table \ref{tab:optimizations21:createrhs1} but for the entire timeloop.}
\label{tab:optimizations21:timeloop1}
\end{table*}
\end{landscape}
\begin{landscape}
\begin{table*}\centering
\scriptsize
\begin{tabular}{c!{\color{black}\vrule}c!{\color{black}\vrule}c!{\color{black}\vrule}c!{\color{black}\vrule}c!{\color{black}\vrule}c!{\color{black}\vrule}cc!{\color{black}\vrule}cc!{\color{black}\vrule}cc!{\color{black}\vrule}ccc}
& & & \bfseries arith. & & & \multicolumn{2}{c|}{\bfseries traffic in GB} & \multicolumn{2}{c|}{\bfseries flops} & \multicolumn{2}{c|}{\bfseries instructions} & \multicolumn{3}{c}{\bfseries loads that hit} \\

& \bfseries runtime & \bfseries \% peak & \bfseries int. & \bfseries \% max. & \bfseries GFlops & \bfseries read & \bfseries write & \bfseries vect. & \bfseries fma & \bfseries mix & \bfseries issue & \bfseries L1P & \bfseries L2 & \bfseries DDR \\
\hline
\bfseries A & 11.8$\,$s & 1.6$\,$\% & 1.7 & 6.4$\,$\% & 37.7 & 15.3 & 6.4 & 31.1$\,$\% & 85.1$\,$\% & 22.7$\,$\% & 31.0$\,$\% & 3.1$\,$\% & 3.9$\,$\% & 0.4$\,$\%\\
\bfseries B & 9.5$\,$s & 1.3$\,$\% & 1.6 & 5.9$\,$\% & 24.9 & 10.5 & 5.6 & 14.4$\,$\% & 80.0$\,$\% & 17.9$\,$\% & 33.2$\,$\% & 1.0$\,$\% & 4.1$\,$\% & 0.3$\,$\%\\
\bfseries C & 4.0$\,$s & 3.0$\,$\% & 0.9 & 24.1$\,$\% & 24.9 & 17.1 & 11.4 & 14.4$\,$\% & 80.0$\,$\% & 17.8$\,$\% & 63.3$\,$\% & 2.4$\,$\% & 8.4$\,$\% & 0.5$\,$\%\\
\bfseries D & 2.2$\,$s & 5.3$\,$\% & 0.9 & 40.9$\,$\% & 24.5 & 16.0 & 11.0 & 33.0$\,$\% & 79.4$\,$\% & 26.0$\,$\% & 56.8$\,$\% & 1.5$\,$\% & 9.6$\,$\% & 0.8$\,$\%\\
\bfseries E & 3.9$\,$s & 3.3$\,$\% & 0.9 & 25.7$\,$\% & 26.0 & 17.5 & 11.8 & 51.0$\,$\% & 85.8$\,$\% & 16.5$\,$\% & 52.5$\,$\% & 1.7$\,$\% & 12.0$\,$\% & 0.7$\,$\%\\
\bfseries F & 2.1$\,$s & 6.2$\,$\% & 0.9 & 51.3$\,$\% & 26.3 & 19.1 & 11.9 & 73.9$\,$\% & 81.3$\,$\% & 20.7$\,$\% & 47.1$\,$\% & 4.4$\,$\% & 13.4$\,$\% & 0.9$\,$\%\\
\hline
\bfseries G & 1.2$\,$s & 11.0$\,$\% & 1.1 & 71.6$\,$\% & 27.0 & 15.1 & 9.7 & 86.4$\,$\% & 75.7$\,$\% & 26.3$\,$\% & 51.7$\,$\% & 6.7$\,$\% & 8.1$\,$\% & 0.9$\,$\%\\
\bfseries H & 1.0$\,$s & 13.3$\,$\% & 1.2 & 78.1$\,$\% & 27.0 & 13.7 & 9.0 & 98.6$\,$\% & 75.7$\,$\% & 28.1$\,$\% & 41.8$\,$\% & 6.0$\,$\% & 13.2$\,$\% & 1.1$\,$\%\\
\bfseries I & 1.1$\,$s & 11.7$\,$\% & 1.2 & 68.4$\,$\% & 27.0 & 14.6 & 7.7 & 98.6$\,$\% & 75.7$\,$\% & 27.1$\,$\% & 40.2$\,$\% & 6.3$\,$\% & 16.7$\,$\% & 0.5$\,$\%\\
\hline
\bfseries T & 0.7$\,$s & 18.9$\,$\% & 1.4 & 100.0$\,$\% & 27.0 & 18.1 & 1.7 & 100.0$\,$\% & & 50.0$\,$\% & 100.0$\,$\% & 0.0$\,$\% & 0.0$\,$\% & 0.0$\,$\%  \\
\bfseries O & 0.5$\,$s & 34.3$\,$\% & 2.5 & 100.0$\,$\% & 33.6 & 11.9 & 1.7 & 100.0$\,$\% & & 50.0$\,$\% & 100.0$\,$\% & 0.0$\,$\% & 0.0$\,$\% & 0.0$\,$\%  \\
	\arrayrulecolor{black}
\end{tabular}
\vskip 0.2cm
\caption{Performance measurements for \textit{create\_rhs} like in Table \ref{tab:optimizations21:createrhs1} but for the baroclinic instability test case for the setup explained in the caption of Table \ref{tab:perfModelCase61}.}
\label{tab:optimizations61:createrhs1}
\end{table*}
\begin{table*}\centering
\scriptsize
\begin{tabular}{c!{\color{black}\vrule}c!{\color{black}\vrule}c!{\color{black}\vrule}c!{\color{black}\vrule}c!{\color{black}\vrule}c!{\color{black}\vrule}cc!{\color{black}\vrule}cc!{\color{black}\vrule}cc!{\color{black}\vrule}ccc}
& & & \bfseries arith. & & & \multicolumn{2}{c|}{\bfseries traffic in GB} & \multicolumn{2}{c|}{\bfseries flops} & \multicolumn{2}{c|}{\bfseries instructions} & \multicolumn{3}{c}{\bfseries loads that hit} \\

& \bfseries runtime & \bfseries \% peak & \bfseries int. & \bfseries \% max. & \bfseries GFlops & \bfseries read & \bfseries write & \bfseries vect. & \bfseries fma & \bfseries mix & \bfseries issue & \bfseries L1P & \bfseries L2 & \bfseries DDR \\
\hline
\bfseries A & 28.0$\,$s & 1.8$\,$\% & 1.0 & 12.4$\,$\% & 101.4 & 64.1 & 34.6 & 22.9$\,$\% & 80.9$\,$\% & 24.6$\,$\% & 32.5$\,$\% & 3.6$\,$\% & 3.6$\,$\% & 0.6$\,$\%\\
\bfseries B & 19.3$\,$s & 1.5$\,$\% & 1.1 & 10.2$\,$\% & 59.8 & 38.0 & 18.1 & 12.6$\,$\% & 77.3$\,$\% & 20.4$\,$\% & 33.5$\,$\% & 2.2$\,$\% & 3.9$\,$\% & 0.5$\,$\%\\
\bfseries C & 9.4$\,$s & 3.4$\,$\% & 0.7 & 37.8$\,$\% & 66.3 & 66.7 & 35.0 & 12.6$\,$\% & 77.3$\,$\% & 19.5$\,$\% & 64.1$\,$\% & 3.2$\,$\% & 7.1$\,$\% & 0.8$\,$\%\\
\bfseries D & 7.6$\,$s & 4.2$\,$\% & 0.7 & 45.3$\,$\% & 65.9 & 64.2 & 34.2 & 20.2$\,$\% & 77.1$\,$\% & 22.4$\,$\% & 62.2$\,$\% & 3.1$\,$\% & 7.3$\,$\% & 1.0$\,$\%\\
\bfseries E & 9.4$\,$s & 3.5$\,$\% & 0.7 & 37.7$\,$\% & 67.3 & 65.9 & 35.2 & 28.1$\,$\% & 79.8$\,$\% & 18.9$\,$\% & 59.9$\,$\% & 3.0$\,$\% & 8.4$\,$\% & 0.9$\,$\%\\
\bfseries F & 7.5$\,$s & 4.4$\,$\% & 0.6 & 49.4$\,$\% & 67.6 & 71.1 & 34.6 & 38.0$\,$\% & 77.9$\,$\% & 20.7$\,$\% & 59.4$\,$\% & 4.0$\,$\% & 7.9$\,$\% & 1.0$\,$\%\\
\hline
\bfseries G & 6.6$\,$s & 5.1$\,$\% & 0.7 & 54.5$\,$\% & 68.3 & 68.9 & 33.7 & 43.9$\,$\% & 75.6$\,$\% & 21.8$\,$\% & 60.8$\,$\% & 4.5$\,$\% & 6.5$\,$\% & 1.0$\,$\%\\
\bfseries H & 6.4$\,$s & 5.2$\,$\% & 0.7 & 55.6$\,$\% & 68.3 & 68.5 & 33.0 & 49.1$\,$\% & 75.6$\,$\% & 21.7$\,$\% & 59.8$\,$\% & 4.2$\,$\% & 7.2$\,$\% & 1.0$\,$\%\\
\bfseries I & 5.8$\,$s & 5.2$\,$\% & 0.9 & 42.1$\,$\% & 62.0 & 47.8 & 22.4 & 48.4$\,$\% & 74.4$\,$\% & 22.1$\,$\% & 55.1$\,$\% & 3.8$\,$\% & 7.8$\,$\% & 0.8$\,$\%\\
\hline
\bfseries T & 3.0$\,$s & 10.2$\,$\% & 0.7 & 100.0$\,$\% & 62.0 & 62.2 & 22.5 & 100.0$\,$\% & & 50.0$\,$\% & 100.0$\,$\% & 0.0$\,$\% & 0.0$\,$\% & 0.0$\,$\%  \\
\bfseries O & 2.2$\,$s & 15.6$\,$\% & 1.1 & 100.0$\,$\% & 68.6 & 44.0 & 17.3 & 100.0$\,$\% & & 50.0$\,$\% & 100.0$\,$\% & 0.0$\,$\% & 0.0$\,$\% & 0.0$\,$\%  \\
	\arrayrulecolor{black}
\end{tabular}
\vskip 0.2cm
\caption{Performance measurements for the entire timeloop like in Table \ref{tab:optimizations21:timeloop1} but for the baroclinic instability test case.\protect\footnotemark[5]}
\label{tab:optimizations61:timeloop1}
\end{table*}
\footnotetext[5]{Even though only \textit{create\_rhs} is changed between the different simulations in this table the runtime saved for the entire timeloop is larger than the time saved in \textit{create\_rhs}. The reason for this behavior is that the reduced runtime of \textit{create\_rhs} leads to an improved synchronization between different MPI processes which reduces the time spent in MPI communication.}
\end{landscape}

To improve vectorization we changed our code in such a way that the operations are performed for the entire element at once (code example \ref{lst:optA}). Our measurements for version F show that this simple modification leads to a significant improvement of the vectorization.

 


\begin{code}
\begin{lstlisting}
real :: rho, rho_x, rho_y, rho_z, u, v, w, rhs
do e=1,num_elem ! loop through all elements
   do i=1,num_points_e ! loop through all points of the element e
      ... ! compute derivatives rho_x, rho_y, rho_z
      rhs = u*rho_x + v*rho_y + w*rho_z + ...
   end do !i
end do !e
\end{lstlisting}
\caption{Fortran code similar to a function from the non-optimized initial version of NUMA (used in versions A, B, C, D and E in Table \ref{tab:optimizations21:createrhs1} to \ref{tab:optimizations61:timeloop1}).}
\label{lst:noopt}
\end{code}

\begin{code}
\begin{lstlisting}
real, dimension(num_points_e) :: rho, rho_x, rho_y, rho_z, u, v, w, rhs
do e=1,num_elem ! loop through all elements
   ... ! compute derivatives like rho_x, rho_y, rho_z
   rhs = u*rho_x + v*rho_y + w*rho_z + ...
end do !e
\end{lstlisting}
\caption{Like code example \ref{lst:noopt} rewritten for improved compiler vectorization (used in version F)}
\label{lst:optA}
\end{code}

\subsection{BG/Q Vector Intrinsics}

To make even better use of the vector unit we rewrote our function \textit{create\_rhs} by using BG/Q vector intrinsics (code example \ref{lst:optC}). We first kept the computation of the derivatives unchanged (version G). This gave us another significant speedup. Using vector intrinsics for the entire function \textit{create\_rhs} gave us another minor speedup (version H). This brings \textit{create\_rhs} to an excellent level of about 80\% of the maximum attainable performance according to the roofline model (column ``\% max.'' in Table \ref{tab:optimizations21:createrhs1} and \ref{tab:optimizations61:createrhs1} and Fig. \ref{fig:roofline21} and \ref{fig:roofline61}).

The fairly large number of loads that hit L1P buffer and L2 cache seems to be due to the random memory access that CG storage produces. Optimal would be if the prefetcher could bring all data into L1 cache before it is needed. We tried different prefetching strategies and handwritten prefetching but could not improve the performance compared to the default strategy. According to our performance model we still expect CG storage to be significantly faster compared to DG storage even though CG storage makes prefetching very difficult.

\begin{code}
\begin{lstlisting}[
	basicstyle=\scriptsize\ttfamily,
]
real, dimension(4,4,4) :: rho, rho_x, rho_y, &
   rho_z, u, v, w, u_x, v_y, w_z, rhs
!IBM* align(32, rho, rho_x, rho_y, rho_z, u, v, w, u_x, v_y, w_z, rhs)
! declare variables representing registers: (each contains four double precision floating point numbers)
vector(real(8)) vct_rho, vct_rhox, vct_rhoy, vct_rhoz
vector(real(8)) vct_u, vct_v, vct_w, vct_rhs
if (iand(loc(rho), z'1F') .ne. 0) stop 'rho is not aligned'
... ! check alignment of other variables
do e=1,num_elem ! loop through all elements
   do k=1,4 ! loop over points in z-direction
      do j=1,4 ! loop over points in y-direction
         ... ! compute derivatives rho_x, ...
         ! load always four floating point numbers:
         vct_u = vec_ld(0, u(1,j,k))
         vct_v = vec_ld(0, v(1,j,k))
         vct_w = vec_ld(0, w(1,j,k))
         vct_rhox = vec_ld(0, rho_x(1,j,k))
         vct_rhoy = vec_ld(0, rho_y(1,j,k))
         vct_rhoz = vec_ld(0, rho_z(1,j,k))
         ! rhs = u*rho_x
         vct_rhs = vec_mul(vct_u,vct_rhox)
         ! rhs = rhs + v*rho_y
         vct_rhs = vec_madd(vct_v,vct_rhoy,vct_rhs)
         ! rhs = rhs + w*rho_z
         vct_rhs = vec_madd(vct_w,vct_rhoz,vct_rhs)
         ! write result from register into cache:
         call vec_st(vct_rhs, 0, rhs(1,j,k))
         ...
      end do !j
   end do !k
end do !e
\end{lstlisting}
\caption{Like code example \ref{lst:noopt} rewritten with vector intrinsics (used in versions G, H and I)}
\label{lst:optC}
\end{code}



\subsection{OpenMP}

OpenMP allows a reduction in the number of MPI processes. This leads for CG storage to a reduced amount of work for some parts of the code (namely the black text in code example \ref{lst:struct}). However, we need to be very careful to avoid race conditions. In \textit{create\_rhs}, race conditions can occur in the summation over all the elements in Eq. (\ref{eq:discstrongform}). Using OpenMP atomic statements made our code too slow. The best solution that we could find was to reorder the elements inside each MPI process in such a way that different OpenMP threads can never compute neighboring elements at the same time. To ensure this we need to synchronize all threads by using an OpenMP barrier after each element computation. These barriers slow down \textit{create\_rhs} by less than 10\% (version I). Nevertheless we obtain in the case of the baroclinic instability a noticeable improvement on the runtime of the entire simulation due to the reduced amount of work for the IMEX corrections in the vertical direction. We obtained the best performance by using 4 OpenMP threads per MPI process (2, 8, 16 and 64 OpenMP threads per MPI process were slower).

Comparing our final version I with our theoretical results (version T) shows reasonably good agreement in the total amount of read and write traffic for the entire timeloop (Table \ref{tab:optimizations21:timeloop1} and \ref{tab:optimizations61:timeloop1}). The measured read and write traffic for \textit{create\_rhs} does not agree very well with our theoretical results. The explanation for this is probably that the L2 cache miss counters that are used to measure the traffic are shared among all hardware threads of the node. Since \textit{create\_rhs} is only a small part of the entire timeloop it can easily happen that our measurements include memory traffic from threads that have not reached \textit{create\_rhs}. This effect would result in too large measurements for the memory traffic. Prefetching on the other hand might lead to data being loaded before the code reaches \textit{create\_rhs} which would result in too low measurements. These effects are not present when measuring the entire timeloop which explains the much better agreement between theoretical results and measurements for the entire timeloop. The number of floating point operations in our theoretical performance model has been tuned to agree with our overall measurements of version I in order to take compiler optimizations into account.

\subsection{Next Steps}

\begin{figure}\centering
\resizebox{9cm}{!}{\includegraphics{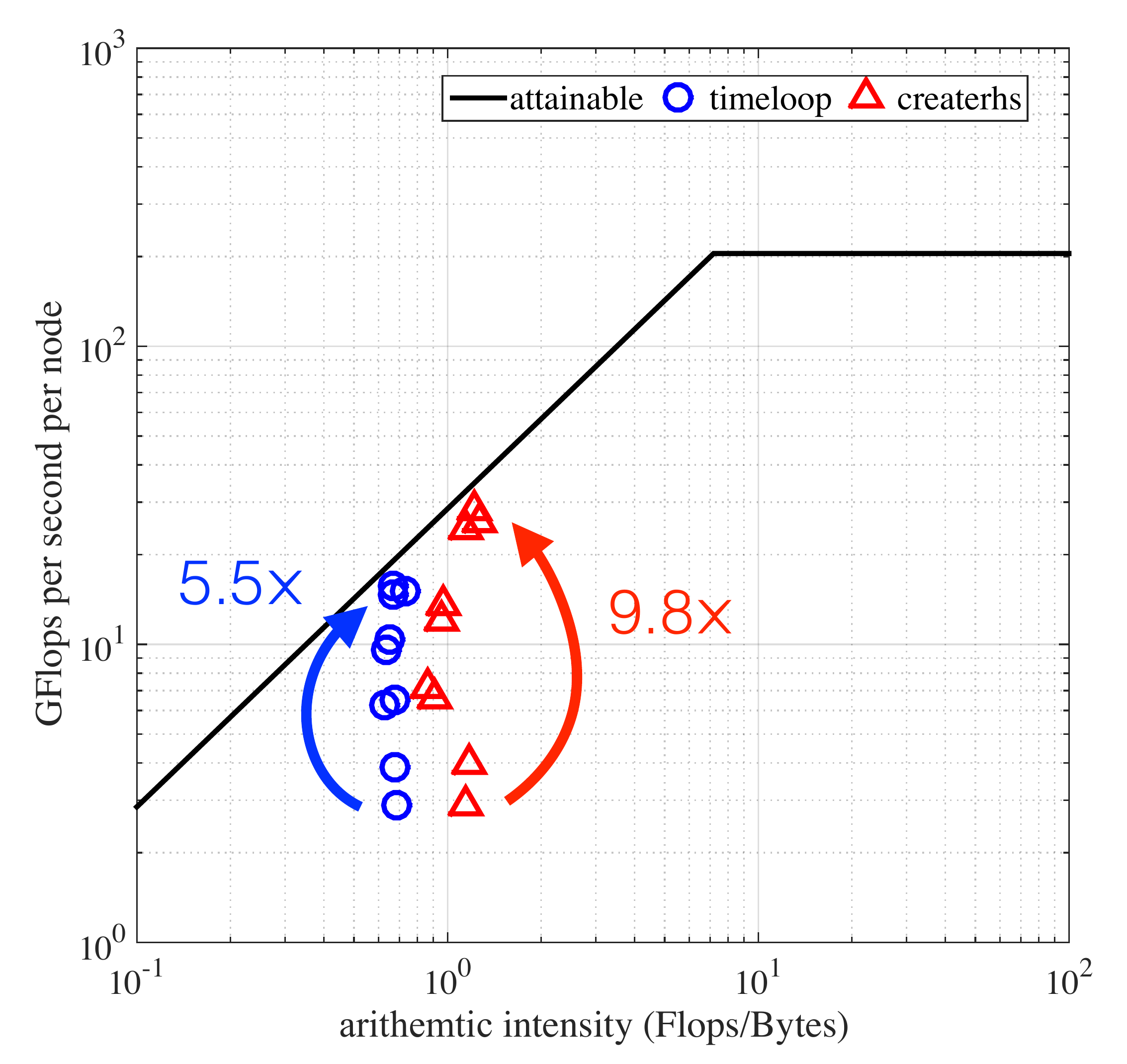}}
\caption{Roofline plot for the different versions shown in Table \ref{tab:optimizations21:createrhs1} and \ref{tab:optimizations21:timeloop1} for the rising thermal bubble test case. For the identification of the different versions we refer to the data in the Tables. The arrows illustrate the performance improvement between the slowest and fastest version.}
\label{fig:roofline21}
\end{figure}

\begin{figure}\centering
\resizebox{9cm}{!}{\includegraphics{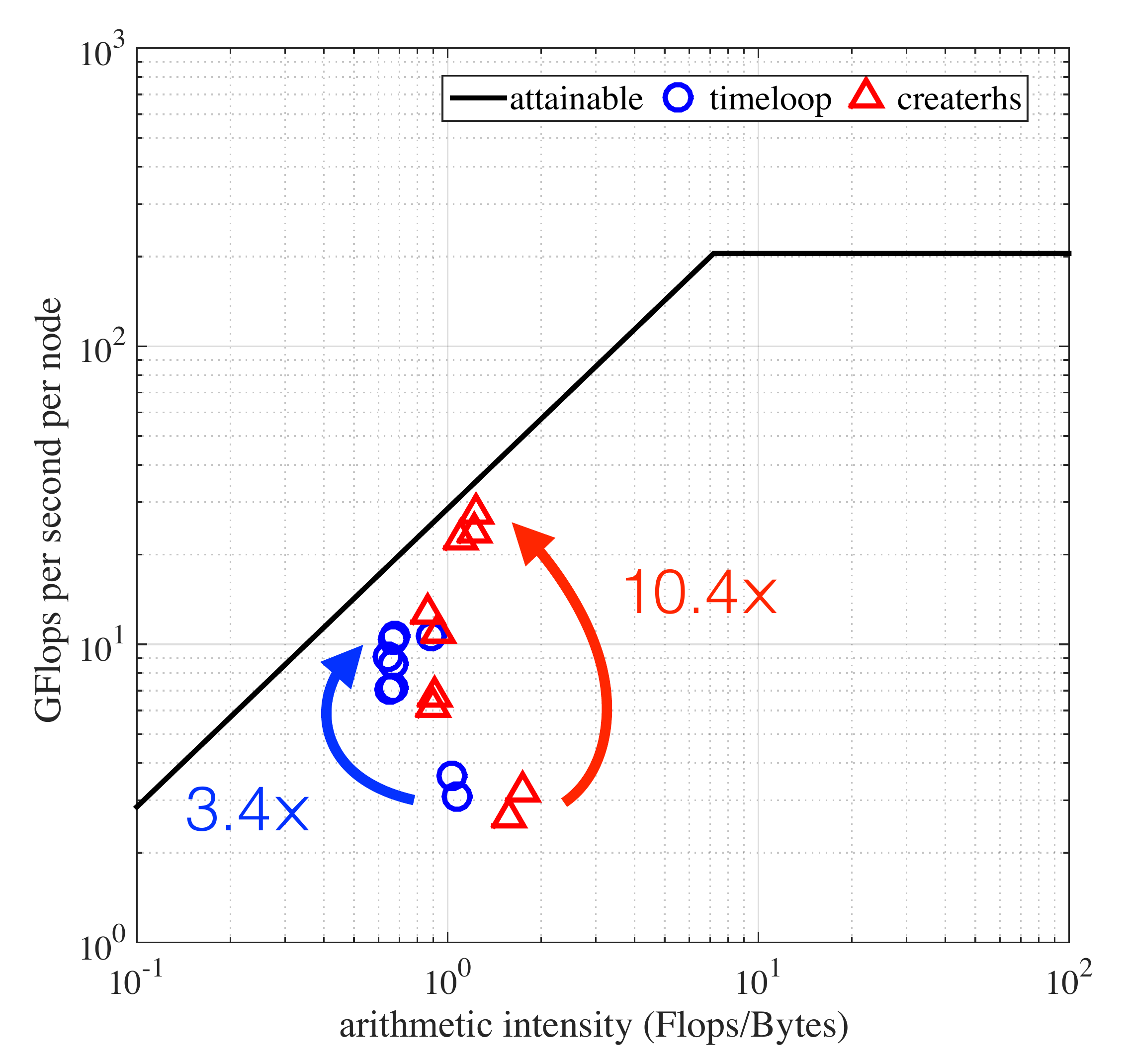}}
\caption{Roofline plot like in Fig. \ref{fig:roofline21} for the different versions shown in Table \ref{tab:optimizations61:createrhs1} and \ref{tab:optimizations61:timeloop1} for the baroclinic instability test case.}
\label{fig:roofline61}
\end{figure}

The roofline plots in Fig. \ref{fig:roofline21} and \ref{fig:roofline61} show that our optimizations have given us a massive improvement in the number of GFlops per second per node which has brought us very close to the maximum attainable performance at the given arithmetic intensity. Our measurements show also that our final version I achieves an excellent level of vectorization (98.6\% of all floating point operations are vectorized). In order to improve performance further we need to increase arithmetic intensity by reducing the overall amount of memory traffic. We have found some unnecessary memory access which we will remove in our future work and we have found that we could reduce memory traffic by recomputing the metric terms in Eq. (\ref{eq:derivatives}) in each stage of our time integration method. These optimizations have been included in our performance model and are shown in version O in Table \ref{tab:optimizations21:createrhs1} to \ref{tab:optimizations61:timeloop1}. According to our performance model we should be able to achieve another factor 2 of speedup for \textit{create\_rhs}. This should bring us to about 35\% of the theoretical peak performance of the processor. Recomputing the metric terms in each time integration stage should also improve the very low percentage of floating point instructions among all instructions (column ``mix'' in Table \ref{tab:optimizations21:createrhs1} to \ref{tab:optimizations61:timeloop1}). We will try these optimizations in our future work.

\section{Strong Scaling Results}
\label{sec:strong_scaling_efficiency}

\begin{figure}\centering
\resizebox{9cm}{!}{\includegraphics{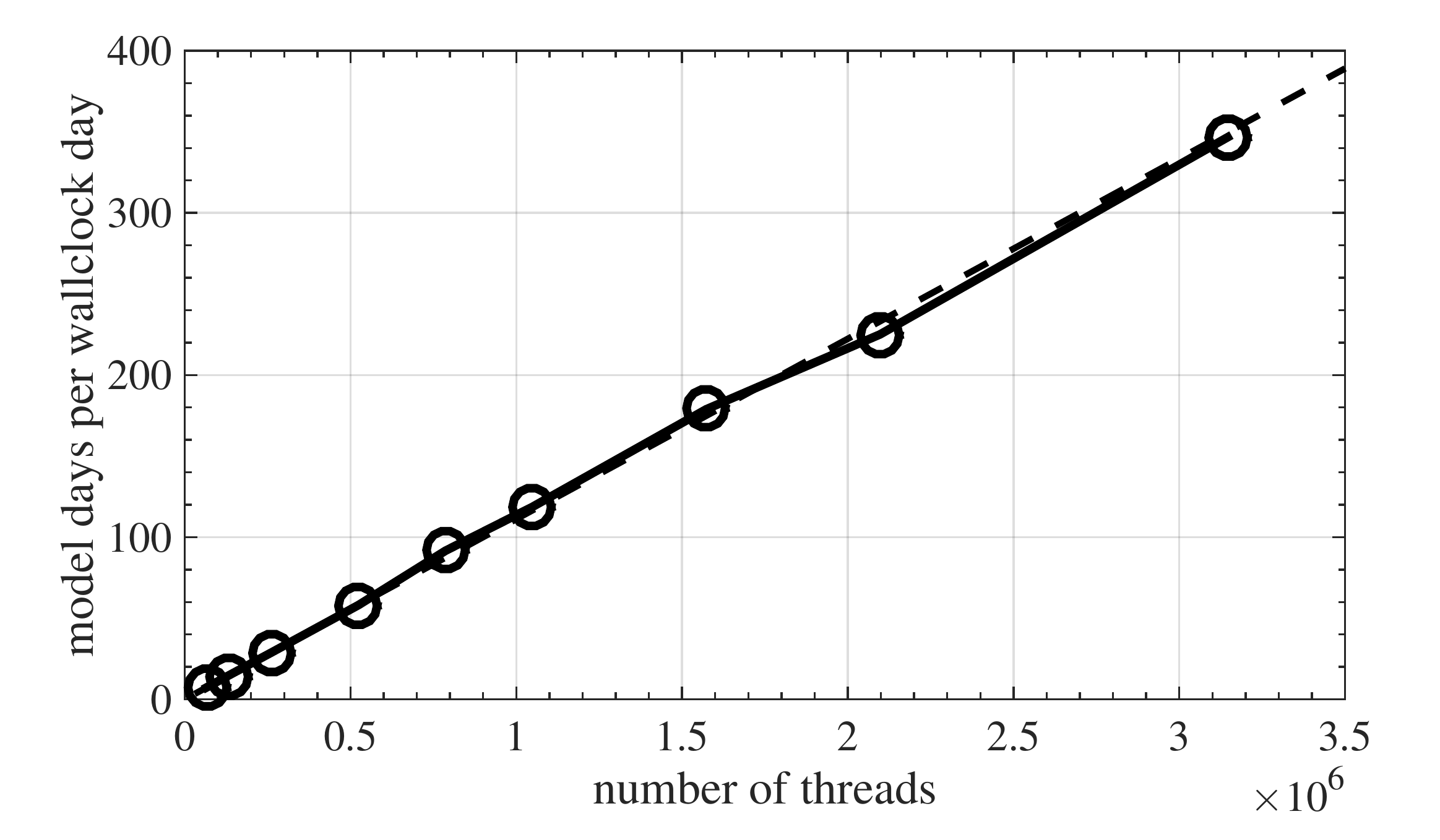}}
\caption{Strong scaling for the baroclinic wave test case with polynomial order ${p=3}$ using 1024 elements per cubed sphere edge and 10 elements in the vertical direction (31 degrees of freedom). This corresponds to a global effective resolution of ${3.0}\,$km and a total number of about 1.8 billion grid points. The dashed line shows ideal strong scaling over a base run on 6.5${\times10^4}$ threads.}
\label{fig:modeldays}
\end{figure}

\begin{figure}\centering
\resizebox{9cm}{!}{\includegraphics{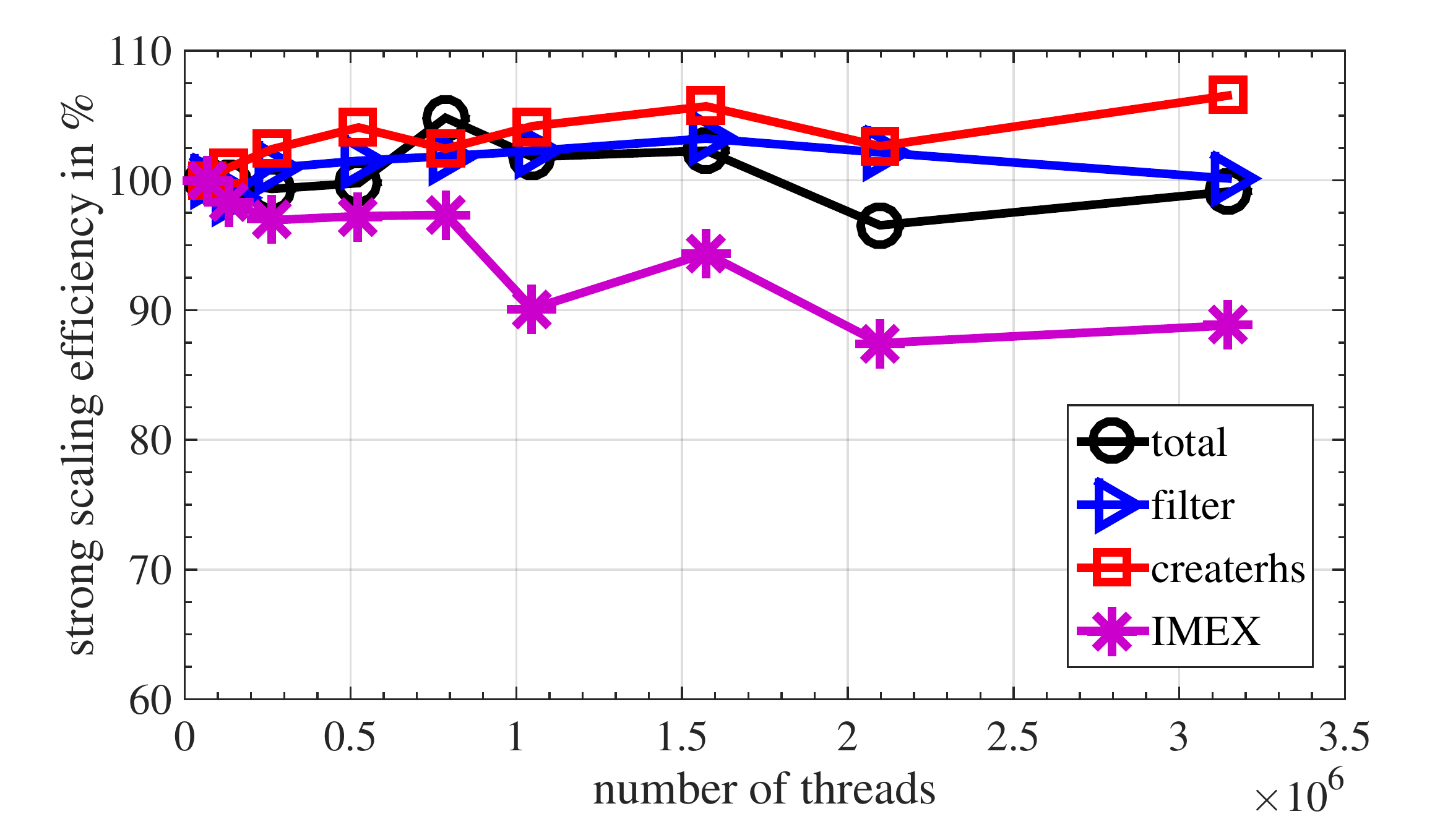}}
\caption{Strong scaling efficiency over base run on 6.5${\times10^4}$ threads for the simulations shown in Fig. \ref{fig:modeldays}. }
\label{fig:barScaling}
\end{figure}

\begin{figure}\centering
\resizebox{9cm}{!}{\includegraphics{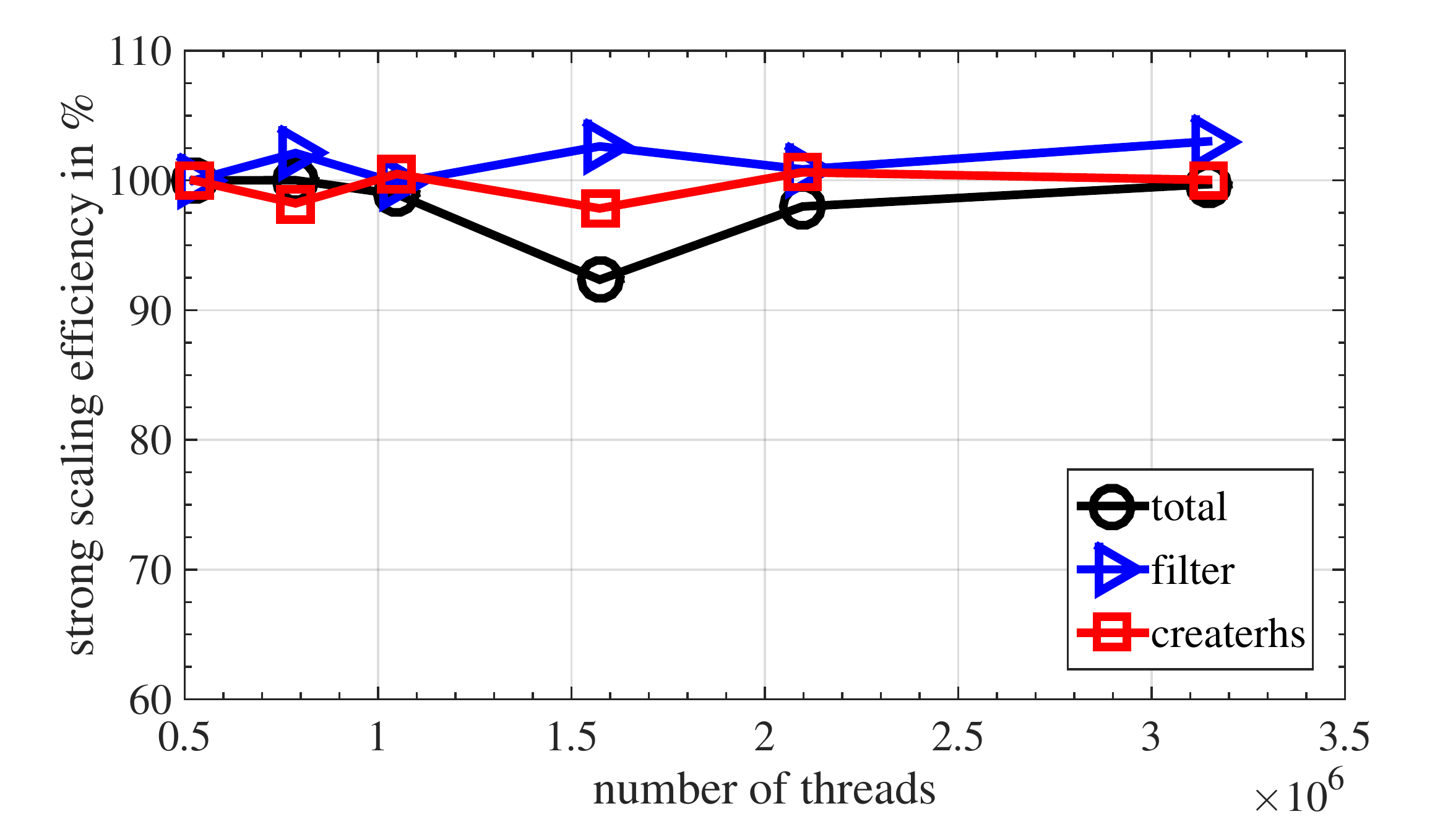}}
\caption{Strong scaling efficiency over base run on 7.7${\times10^5}$ threads for the rising thermal bubble test case using ${1024\times1024\times1536}$ elements which corresponds to about 43 billion grid points.}
\label{fig:rtbScaling}
\end{figure}

\begin{figure}\centering
\resizebox{9cm}{!}{\includegraphics{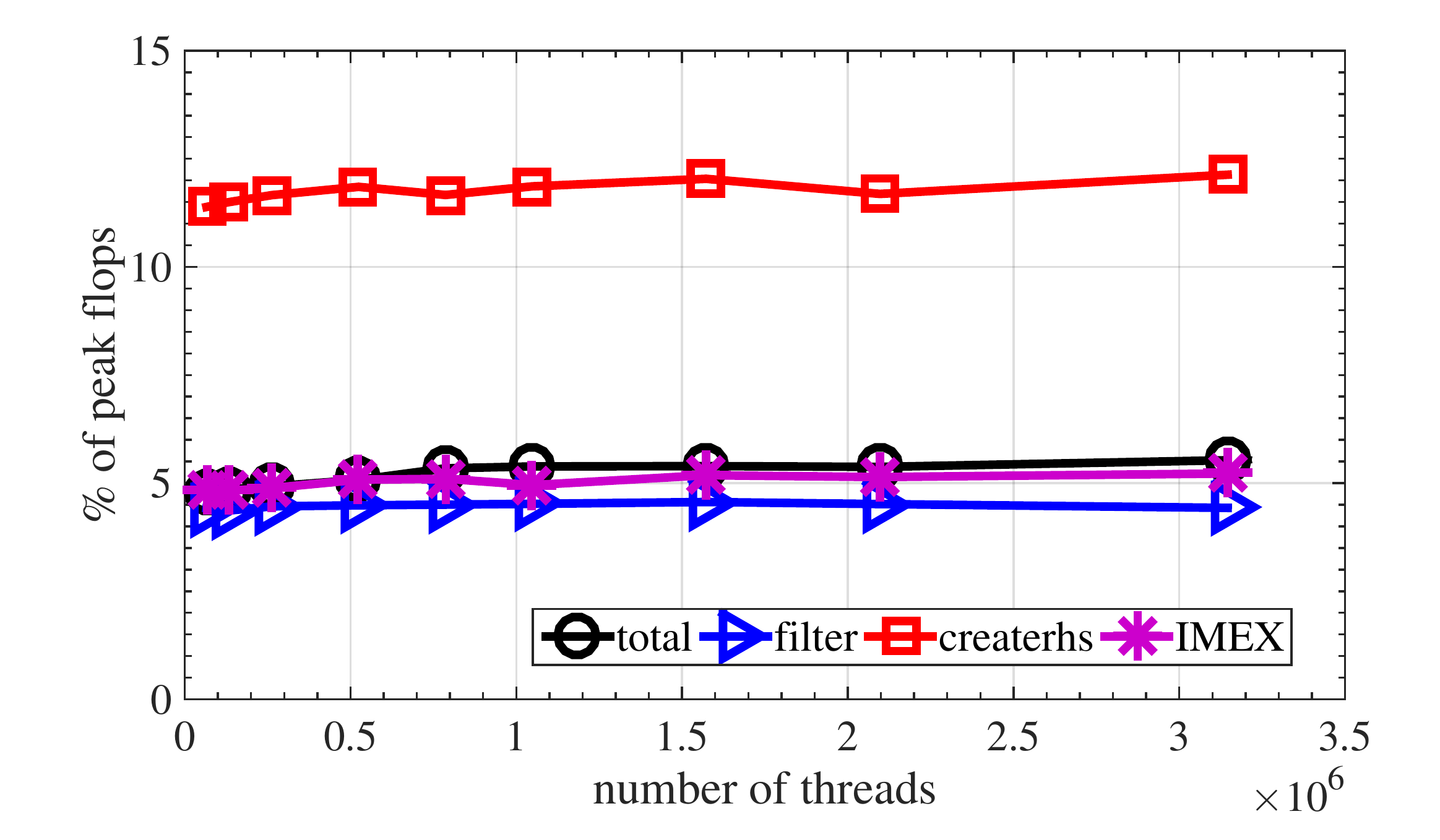}}
\caption{Percentage of theoretical peak performance in terms of floating point operations for the baroclinic wave test case like in Fig. \ref{fig:modeldays}.}
\label{fig:barFlops}
\end{figure}

\begin{figure}\centering
\resizebox{9cm}{!}{\includegraphics{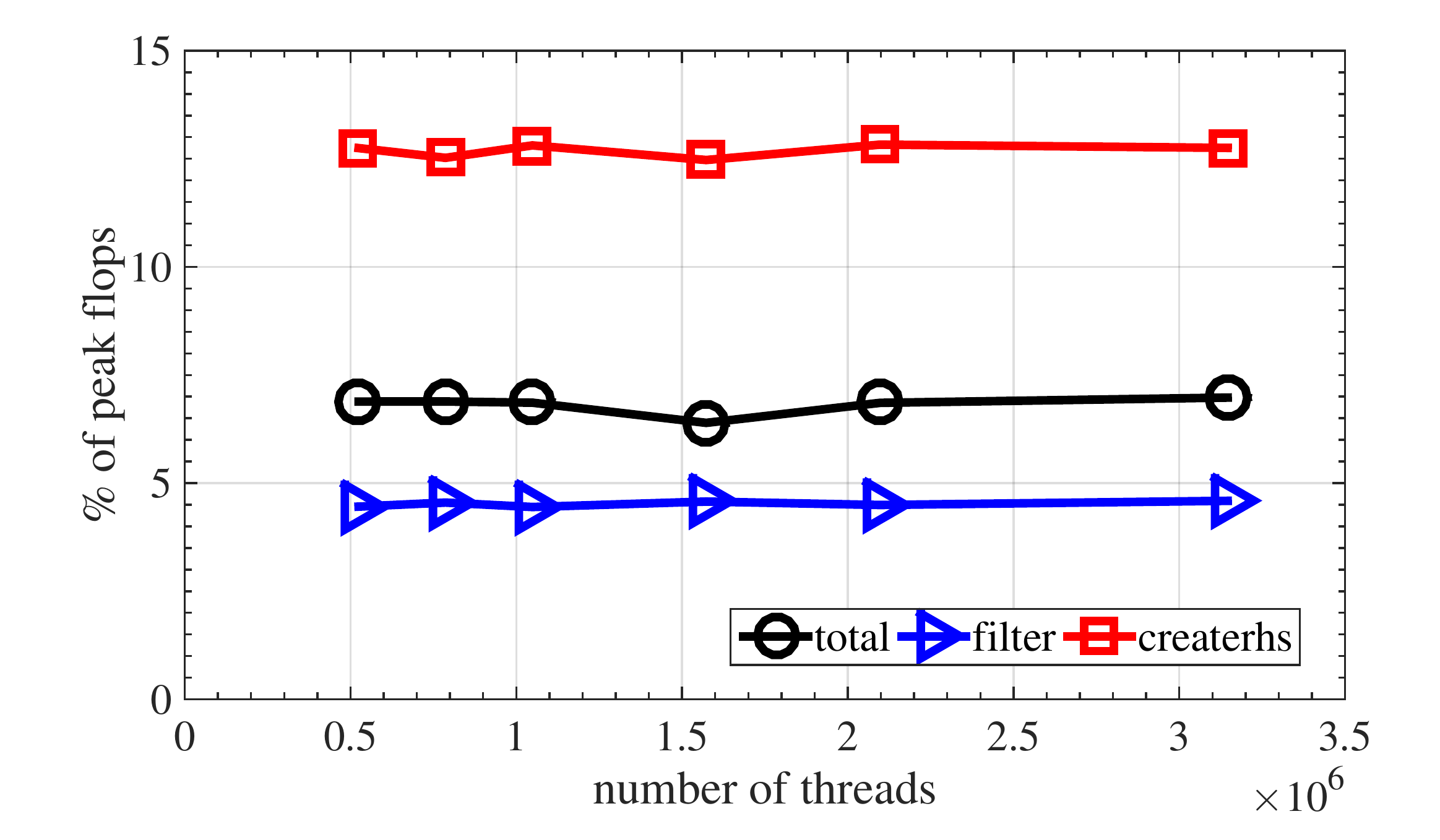}}
\caption{Percentage of theoretical peak performance in terms of floating point operations for the rising thermal bubble test case like in Fig. \ref{fig:rtbScaling}.}
\label{fig:rtbFlops}
\end{figure}

We present in this section strong scaling results up to the entire machine Mira for the baroclinic wave test case (Fig. \ref{fig:modeldays}, \ref{fig:barScaling} and \ref{fig:barFlops}) and the rising thermal bubble test case (Fig. \ref{fig:rtbScaling} and \ref{fig:rtbFlops}). All these results use version I from Section \ref{sec:optimizations}.

The runtime of the entire simulation for the baroclinic wave test case is shown in Fig. \ref{fig:modeldays}. The dynamics of a one day forecast needs to be finished within less than about 4.5 minutes runtime (more than 320 model days per wallclock day). We reach this goal on the entire machine Mira for our 3.0$\,$km uniform horizontal resolution simulation of the baroclinic wave test case which takes 4.15 minutes runtime per one day forecast (346.6 model days per wallclock day).

The strong scaling efficiency of the simulations in Fig. \ref{fig:modeldays} is shown in \mbox{Fig. \ref{fig:barScaling}} for the different parts of the code. The entire code reaches a strong scaling efficiency of 99.1\% on the entire machine Mira. The parts \textit{create\_rhs} and filter show a scaling efficiency of more than 100\%. This is not surprising because the problem fits better into L2 cache with increasing number of threads and at the same time the time spent in our OpenMP barriers is decreasing. The IMEX part gives us the lowest scaling efficiency. We still need to understand the reason for this behavior.

The lowest scaling efficiency for the entire simulation is obtained for 2.1$\times10^6$ threads. This is due to non-optimal load balancing. The number of elements per thread is always perfectly balanced for all results shown in this paper. However, the arrangement of these elements can vary which leads for CG storage to variations in the number of grid points. So far we use the very simple mesh partitioning built into the p4est library. We expect to be able to improve this result by using more advanced mesh partitioning algorithms.

The strong scaling efficiency of the rising thermal bubble test case is shown in Fig. \ref{fig:rtbScaling}. We achieve 99.7\% strong scaling efficiency on the entire machine for this case. We use a much larger total number of grid points of about 43 billion grid points for this case because we plan to use our code for hurricane and cloud simulations at this kind of problem size. We have not optimized the memory usage of our code. The smallest number of threads that can handle this problem is currently 7.7$\times10^5$. We expect to be able to reduce the memory usage of our code significantly. Also we need to understand the reason why the simulation using 1.6$\times10^6$ shows a reduced performance.

The percentage of the theoretical peak performance in terms of floating point operations is shown in Fig. \ref{fig:barFlops} and \ref{fig:rtbFlops}. Not surprisingly we obtain the best performance for our optimized part \textit{create\_rhs}. For the baroclinic wave test \textit{create\_rhs} reaches 1.21 PFlops (12.1\% of peak)  and for the rising thermal bubble test it reaches 1.28 PFlops (12.8\% of peak) on the entire machine. The sustained performance of the entire simulation is at 0.55 PFlops for the baroclinic wave test and at 0.70 PFlops for the rising thermal bubble test on the entire machine Mira.

\section{Conclusions}
\label{sec:conclusions}

In this paper, we present the optimization and performance analysis of the atmospheric model NUMA. Our optimizations have improved the performance of our code by a factor of about 10 and have brought us very close to the maximum attainable performance due to the peak memory bandwidth (Fig. \ref{fig:roofline21} and \ref{fig:roofline61}). These optimizations allow us to perform most of the computations at 1.2 PFlops on the entire supercomputer Mira by using BG/Q vector intrinsics. The sustained performance of the entire simulation is at 0.70 PFlops for a rising thermal bubble test case using explicit time integration. We have not optimized all parts of the code yet. For the baroclinic wave test the non-optimized computations for the implicit part of the time integration lead to a slightly lower sustained performance of 0.55 PFlops. We expect to improve our performance significantly by optimizing the remaining non-optimized parts of our code.

We have shown that NUMA achieves a near perfect strong scaling efficiency of 99.7\% for the rising thermal bubble test case using 43 billion grid points on the entire 3.14 million threads of Mira. For the baroclinic wave test case on the sphere we obtain a strong scaling efficiency of 99.1\% using a mesh with 1.8 billion grid points. This allows us to compute a one day forecast at 3.0$\,$km resolution within 4.15 minutes runtime and fulfills the requirements for operational weather prediction (less than 4.5 minutes runtime for the dynamics of a one day forecast).

As explained in the introduction, we expect this massive increase in resolution to be a major step towards more accurate weather forecasts. Nevertheless, the demand to increase the resolution of NWP models does not end at 3$\,$km resolution \citep{bauer2015quiet}. The demand for better performance is even more severe when high resolution climate prediction is considered. Climate prediction requires forecast periods of more than one hundred years. To simulate such a long period of time at a resolution of 3$\,$km would still require about one year of runtime on the entire machine Mira when tracers and physics parameterizations are taken into account. For this reason we need to continue to work on improving the performance of our code and to optimize it for next generation supercomputers.

Our analysis in this paper shows that we need to reduce the amount of memory traffic to further improve our performance. It should be possible to achieve this by recomputing metric terms in each stage of our time integration method. For this reason, we expect to improve our sustained performance of the entire simulation beyond 1 PFlops. This should allow us to reach a uniform horizontal resolution close to 2$\,$km within operational requirements. The next goal will be the optimization of our code for the upcoming next generation supercomputer Aurora at the Argonne National Laboratory. Aurora is expected to achieve a peak performance of 180 PFlops (18 times more than Mira). We hope to be able to reach 1$\,$km resolution for global numerical weather prediction once Aurora is available and once our code is fully optimized for that machine.

\begin{acks}
\label{sec:acknowledgement}
This research used resources of the Argonne Leadership Computing Facility, which is a DOE Office of Science User Facility supported under Contract DE-AC02-06CH11357. We would like to thank Vitali Morozov at the Argonne National Laboratory for his support in analyzing the performance of our code with the Hardware Performance Monitor Toolkit. AM, MK, and SM are grateful to the National Research Council of the National Academies.
\end{acks}

\begin{dci}
The Authors declare that there is no conflict of interest.
\end{dci}

\begin{funding}
This work was supported by the Office of Naval Research [PE-0602435N]; the Air Force Office of Scientific Research (Computational Mathematics program); and the National Science Foundation (Division of Mathematical Sciences) [121670].
\end{funding}



\bibliographystyle{sageh}
\bibliography{am}

\end{document}